\documentclass[twocolumn]{article}

\usepackage[letterpaper,textwidth=6.75in,textheight=9in,left=0.77in,top=0.9314in,headheight=10pt,headsep=10pt,footskip=25pt,columnsep=0.25in]{geometry}
\usepackage[T1]{fontenc}
\usepackage{times}
\usepackage[numbers,sort&compress,square]{natbib}
\setcitestyle{citesep={,}}
\usepackage{fancyhdr}
\usepackage{caption}
\usepackage{microtype}
\usepackage{graphicx}
\usepackage{booktabs}
\usepackage{amsmath,amssymb}
\usepackage{colortbl}
\usepackage[most]{tcolorbox}
\usepackage{enumitem}
\usepackage{xcolor}
\usepackage{tabularx}
\usepackage{hyperref}
\definecolor{verdictyes}{RGB}{0,150,50}
\definecolor{verdictweak}{RGB}{160,140,40}
\definecolor{cntopt}{RGB}{128,45,165}
\definecolor{cntsim}{RGB}{205,95,0}
\definecolor{cntmatch}{RGB}{0,135,60}
\definecolor{cntperop}{RGB}{200,20,20}
\definecolor{cntfree}{RGB}{110,110,110}
\definecolor{verdictno}{RGB}{125,125,125}
\newcommand{\lH}{\textcolor{cntopt}{\textbf{H}}}
\newcommand{\lS}{\textcolor{cntsim}{\textbf{S}}}
\newcommand{\lD}{\textcolor{cntmatch}{\textbf{D}}}
\newcommand{\lF}{\textcolor{cntfree}{\textbf{F}}}
\newcommand{\lP}{\textcolor{cntperop}{\textbf{P}}}
\newcommand{\lN}{\textcolor{black}{\textbf{N}}}
\hypersetup{colorlinks=true,linkcolor=purple,citecolor=cyan,urlcolor=magenta,
  pdftitle={Optical Neural Networks Should Be Evaluated by Function, Not Only Physics},
  pdfauthor={Shi-Yuan Ma, Xingjian Bai, Ryan Hamerly, Mahmoud Jalali Mehrabad}}

\makeatletter
\renewcommand\small{\@setfontsize\small\@ixpt{10}}
\renewcommand\footnotesize{\@setfontsize\footnotesize\@ixpt{10}}
\renewcommand\scriptsize{\@setfontsize\scriptsize\@viipt{8}}
\renewcommand\Large{\@setfontsize\Large\@xivpt{16}}
\renewcommand\@seccntformat[1]{\csname the#1\endcsname.~}
\renewcommand\section{\@startsection{section}{1}{\z@}{-0.12in}{0.02in}{\large\bfseries\raggedright}}
\renewcommand\subsection{\@startsection{subsection}{2}{\z@}{-0.10in}{0.01in}{\normalsize\bfseries\raggedright}}
\renewcommand\subsubsection{\@startsection{subsubsection}{3}{\z@}{-0.08in}{0.01in}{\normalsize\scshape\raggedright}}
\renewcommand\paragraph{\@startsection{paragraph}{4}{\z@}{1.5ex \@plus 0.5ex \@minus 0.2ex}{-1em}{\normalsize\bfseries}}
\renewcommand\footnoterule{\kern-3pt\hrule width 0.8in\kern 2.6pt}
\def\@listii{\leftmargin\leftmarginii\labelwidth\leftmarginii\advance\labelwidth-\labelsep
  \topsep 2pt plus 1pt minus 0.5pt\parsep 1pt plus 0.5pt minus 0.5pt\itemsep\parsep}
\makeatother
\renewenvironment{abstract}{\centerline{\large\bfseries Abstract}\vspace{-0.12in}\begin{quote}}{\par\end{quote}\vskip 0.12in}
\definecolor{DefFrame}{HTML}{FED739}
\tcbset{
  defboxstyle/.style={
    enhanced, colback=white, colframe=DefFrame, boxrule=2pt, arc=3.5mm, boxsep=0mm,
    left=4mm,right=4mm,top=3mm,bottom=3mm, fonttitle=\large\bfseries, coltitle=black,
    boxed title style={colback=white, boxrule=0pt, sharp corners, left=0pt,right=0pt,top=0pt,bottom=0pt},
    attach boxed title to top left={xshift=3mm, yshift=-2.3mm},
  }
}
\newtcolorbox{DefinitionBox}[2][]{defboxstyle, title={#2}, #1}
\newtcolorbox{expressivitycard}{
    enhanced, colback=gray!5, colframe=black!70, arc=1mm, boxrule=0.7pt,
    left=3mm, right=3mm, top=1mm, bottom=1mm,
    fonttitle=\bfseries\color{black}, title=Expressivity Card, attach title to upper,
}

\newcommand{\R}{\mathbb{R}}
\begin{document}

\twocolumn[
\thispagestyle{plain}
{\centering
\baselineskip 18pt
\hrule height 1pt
\vskip 0.25in
{\Large\bfseries Optical Neural Networks Should Be Evaluated\\by Function, Not Only Physics}
\vskip 0.22in
\hrule height 1pt
\vskip 0.3in
\textbf{Shi-Yuan Ma}\,\textsuperscript{1}\enskip
\textbf{Xingjian Bai}\,\textsuperscript{1}\enskip
\textbf{Ryan Hamerly}\,\textsuperscript{1,2}\enskip
\textbf{Mahmoud Jalali Mehrabad}\,\textsuperscript{1}\par}
\vskip 0.3in
]

{\renewcommand{\thefootnote}{}\footnotetext{\hspace*{-\footnotesep}\textsuperscript{1}Massachusetts Institute of Technology, Cambridge, MA, USA. \textsuperscript{2} NTT Research Inc., Sunnyvale, CA, USA. Correspondence to: Shi-Yuan Ma \textless{}\href{mailto:mashiyua@mit.edu}{mashiyua@mit.edu}\textgreater{}.}}

\begin{abstract}
Optical neural networks (ONNs) are attracting significant attention as alternative platforms for machine learning (ML). In an ONN, an optical module often takes over part of the computation inside a digital ML pipeline. \emph{How much of the task does the optical module itself perform, and how much does it contribute to any claimed system-level advantages, such as throughput or energy?}
This paper offers a simple, function-oriented evaluation methodology. Whatever its internal physics, the optical module appears to the surrounding digital pipeline as a mapping between real-valued tensors. We therefore evaluate it in three steps: (i) define its boundary in the pipeline, as real-valued tensors in and out; (ii) replace it with standard digital modules to identify those that match the optical system's performance on the same task and metric; and (iii) state system-level claims, such as throughput or energy, per inference against the performance-matched digital module. We summarize steps (i) and (ii) in an \emph{Expressivity Card} and demonstrate its use on published ONN systems. This helps isolate the optical module's contribution and provides a concrete digital reference for assessing its claimed system-level advantages, without the need to understand the physics.
While we focus on optical systems as the most prominent example, the same approach applies to any physical analog computing platform used in ML, wherever a physical module takes over part of a digital pipeline.
\end{abstract}

\section{Introduction}
\label{sec:intro}

\begin{figure}[!t]
  \centering
  \includegraphics[width=\columnwidth]{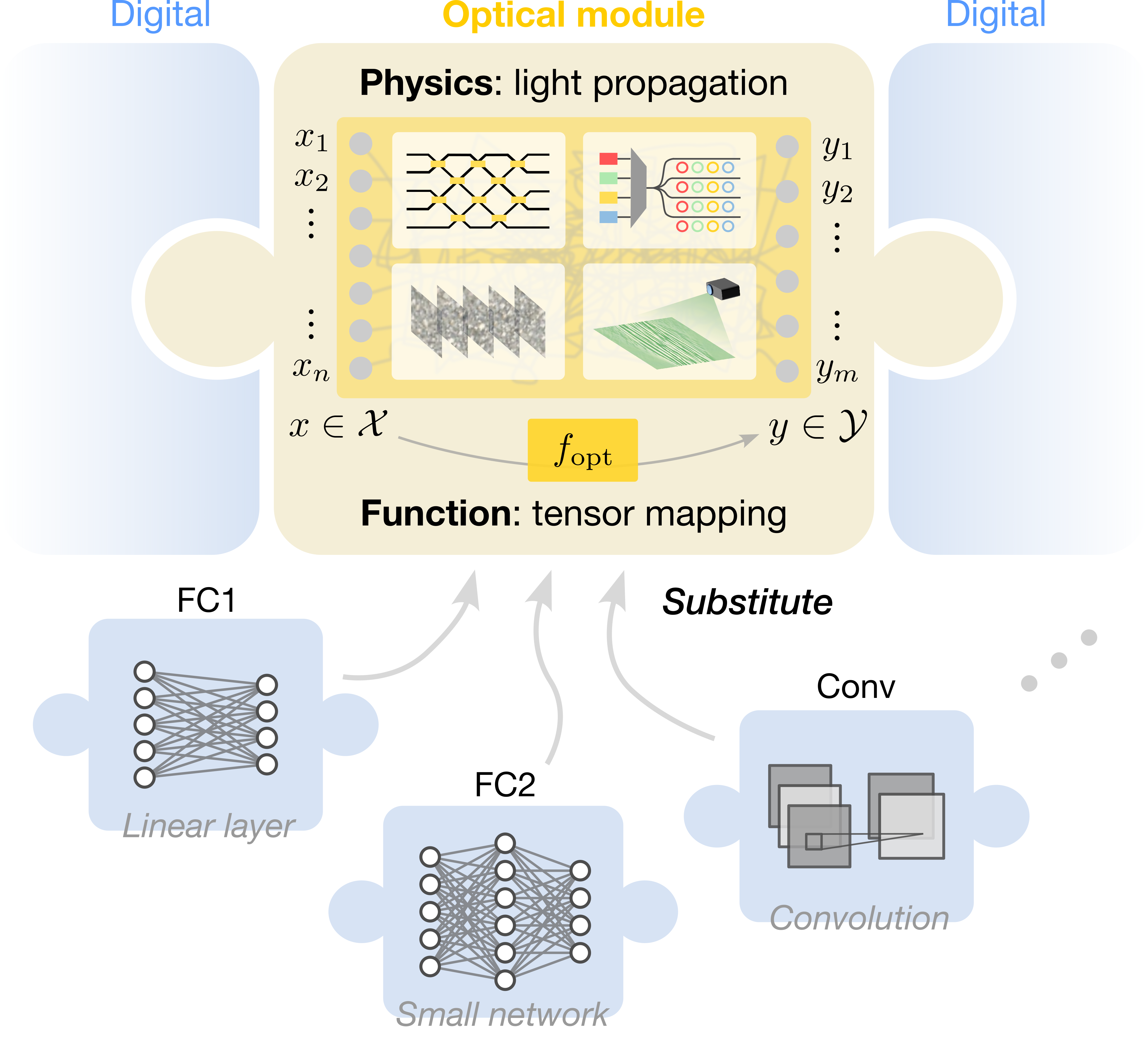}
  \caption{\textbf{Evaluating optical modules by function, not only physics.}
    An optical module (yellow) sits within a hybrid digital--optical pipeline.
    Internally, the module involves complex light propagation through various physical substrates
    (interferometer meshes, wavelength-division multiplexing, diffractive elements, programmable waveguides, etc.).
    However, when used as a component in a classical ML pipeline, the surrounding digital system
    interacts with it only through real-valued tensors: input $x \in \mathcal{X}$ and output $y \in \mathcal{Y}$.
    The mapping $f_{\mathrm{opt}}$ is therefore the relevant object for ML comparison.
    Its empirical expressivity is measured by an ablation study: the optical module is replaced by standard digital modules with the same inputs and outputs, and the one that matches its task performance is identified.}
  \label{fig:teaser}
\end{figure}

Optical and photonic computing platforms---physical systems that use light to perform computation---are rapidly gaining visibility as potential complements to conventional platforms for machine learning (ML) \cite{wetzstein2020inference,shastri2021photonics,mcmahon2023physics,farmakidis2024integrated,savage2026light}. Although the field remains controversial, even skeptics cannot deny its popularity in both industry and academia, as reflected by significant venture investment ($>$\$4 billion raised in startups as of May 2026; see Appendix~\ref{app:industry}) and a growing presence in high-impact journals \cite{zhu2022space,wang2022optical,sludds2022delocalized,pai2023experimentally,chen2023deep,bai2023microcomb,meng2023compact,zhou2023memory,lin2024120,bandyopadhyay2024single,xu2024large,dong2024partial,xue2024fully,hu2024diffractive,hu2025integrated,li2025spatial,ma2025quantum,wang2025diffractive,ahmed2025universal,hua2025integrated,chen2025optical,chen2025all,pintus2025integrated}.
At the same time, the technical conversation remains fragmented: ONN papers often communicate technical advances in physics-native terms (light propagation, diffractive elements, interferometer meshes), while ML audiences reason about modules, interfaces, constraints, and empirical baselines. When these languages do not meet, claims about ONN capabilities become difficult to interpret---not because the underlying engineering is weak, but because the evaluation object is underspecified for the intended ML use case.

Consider an ONN system used for image generation. Despite its name, the system may perform only part of its computation optically, with digital processing around the optical module. The reported image quality reflects the entire hybrid pipeline, not just the optical part. The role of the optical module can be even less apparent in more complex settings, such as reinforcement learning, where the entire hybrid system may be used repeatedly. This raises a natural question: how much does the optical module itself contribute?

The core issue is simple. Any module in an ML pipeline, whether a convolutional layer or an attention block, takes in a tensor of real numbers and returns another tensor of real numbers. When the optical hardware in an ONN system serves as a component in a classical ML pipeline, the same holds: the surrounding computation interacts with it only through real-valued tensors, a control tensor in and a measurement tensor out. Internal physics such as complex optical fields, wave interactions, and resonant dynamics is invisible at this interface.
The resulting \emph{real-to-real mapping} is therefore the relevant object for ML comparison: it is what the optical module does in the pipeline, which is what we mean by its function.\footnote{We use ``function'' in its everyday sense (``what is the function of this part?''): what the module does in the pipeline, not how it is physically realized. We do not mean strict functional equivalence, a digital module reproducing the optical mapping point by point; the comparison is of task performance, as in any ablation study (Sec.~\ref{sec:equivalence}). In general, ONN papers likewise show the value of the optical modules by task performance, not by such equivalence (Sec.~\ref{sec:background}).} When this mapping is left implicit, ML readers cannot assess the optical module's contribution to the overall pipeline.

\vspace{5pt}
\begin{DefinitionBox}{Function-oriented evaluation}
\textbf{ONN systems should clearly define the boundary of the optical module in the digital ML pipeline: what goes in and what comes out, as real-valued tensors. We should then report ablation studies that isolate its contribution, as ML papers do for any new module, so that readers can see it without needing to know the physics.}
\end{DefinitionBox}
\vspace{10pt}

Figure~\ref{fig:teaser} illustrates this perspective. An optical module (yellow) sits within an ML pipeline. While internally it relies on light propagation through various physical substrates, at the digital interface it simply implements a real-valued tensor mapping $f_{\text{opt}}: \mathcal{X} \to \mathcal{Y}$.
The key evaluation question then becomes: which standard digital modules would match the optical module's task performance if put in its place?

In practice, this can come down to three steps:
\begin{enumerate}
    \item \textbf{Define the optical-module boundary in the ML pipeline explicitly:} State what goes into the optical module and what comes out, as real-valued tensors seen by the surrounding digital pipeline: how the input is written into the optics, what is read out, and the sizes and constraints of both.
    \item \textbf{Report empirical expressivity using standard digital substitutes:} Replace the optical module with standard digital substitutes with the same inputs and outputs, and report task performance using the identical metrics applied to the optical module. This is an ablation study, as ML papers routinely report for a new module.
    \item \textbf{Anchor system-level claims to performance-matched substitutes:} State system-level claims, such as speed or energy, per inference against the digital substitute that matches the optical module's task performance. Only then do such claims compare modules that do the same job.
\end{enumerate}

The paper develops as follows. Alongside ONNs whose optics implements a named operation, such as a matrix multiplication, a newer branch of ONN research builds optical modules whose value is shown by task performance alone (Sec.~\ref{sec:background}). For these systems, the optical module's contribution is less obvious. In ML, an ablation study of a proposed new module is routine; in ONN papers it is largely missing (Sec.~\ref{sec:background}, Appendix~\ref{app:census}). 
Section~\ref{sec:motivation} explains why its absence is easy to overlook, and why operation counts from the hardware and system-level claims stated per operation do not provide a shared reference. 
Section~\ref{sec:scope} clarifies the scope of our proposal, and Section~\ref{sec:boundary} introduces the Expressivity Card (EC) and how it anchors system-level claims. 
We then demonstrate the card on existing ONN systems, illustrating how it can be applied and what it reveals about the optics' contribution (Sec.~\ref{sec:casestudies}, Appendices~\ref{app:caseB} and \ref{sec:caseA}). 
Because the paper speaks to both the optics and the ML communities, some of its points will read differently to different readers, and we hope to encourage discussion. Section~\ref{sec:altviews} takes up the alternative views we expect most often, including the role of physics-first descriptions, the task dependence of empirical expressivity, and the cost and fairness of the proposed ablation study.

\section{Background}
\label{sec:background}
\paragraph{From named operations to \texorpdfstring{\textit{task-level}}{task-level} optical modules.}
Historically, a large class of ONN-for-ML work presents optical hardware as a direct implementation of a clearly defined linear-algebra primitive used in standard neural networks; we call these \emph{operation-level} modules. Examples include matrix--vector multiplication \cite{morizur2010programmable,popoff2011controlling,labroille2014efficient,shen2017deep,lin2018all,hamerly2019large,spall2020fully,buddhiraju2021arbitrary,wang2022optical,gigan2022imaging,bernstein2023single,meng2025digital,onodera2025arbitrary}, matrix--matrix multiplication \cite{xu2022high,dong2023higher,wang2025diffractive,zhang2025direct,luan2026single}, and convolution \cite{chang2018hybrid,colburn2019optical,feldmann2021parallel,xu202111,cui2025spectral}.
For this class of work, the mapping implemented by the optical module is relatively clear to ML researchers: it is an (often constrained) linear operator.

Motivated by the desire to go beyond ``linear-only'' primitives, recent ONN research explores mechanisms yielding nonlinear input--output behavior in largely optical forward paths \cite{wanjura2024fully,li2024nonlinear,xia2024nonlinear,yildirim2024nonlinear,wang2024large}.
Following these ideas, several systems report impressive task-level demonstrations beyond linear-only operations in the optical domain \cite{xu2024large,chen2025optical,chen2025all,wang2024large,zhan2024photonic,liu2026nonlinear,ng2026few,fu2023photonic,wang2024ultrafast,aadhi2025scalable,parto2026high,wang2025optical,petrini2026degree,gao2023optofluidic,chen2023all,wang2023image,peng2026optical}. In these systems, the optical path may no longer be advertised as ``just a linear layer.''
Without a named operation to point to, these papers show the value of the optical module the way ML papers show the value of a model: by task performance, such as classification accuracy or the quality of generated images, for example by showing that the optics does better than a linear model on the same task. We call these modules \emph{task-level}. Task performance is thus the measure the authors choose to demonstrate their work, and it is the measure the ablation study we propose uses.
This progress is exciting---it suggests that optical modules can participate in richer computations than operation-level accelerators---but it also obscures what the optical component actually contributes: the representational power is no longer obvious from physics diagrams alone.

\paragraph{The resulting gap: unclear expressivity of the optical module.}
When task performance is reported for a whole pipeline, a reader, especially one outside optics, naturally asks how much of it comes from the optical module itself and how much from the digital parts around it.
Task-level ONN papers necessarily describe internal physical states---complex optical fields, wave interactions, resonant dynamics---using abstractions far from those ML readers use. When the real-to-real mapping is left implicit, an ML reader cannot tell how expressive the optical module actually is. Crucially, this ambiguity also affects ONN researchers: without shared descriptions of what the module does, comparing devices across platforms or assessing whether a new demonstration reflects increased expressivity versus a different surrounding pipeline becomes difficult.

The gap is visible in what these papers report. Their baselines answer real questions, but the ablation study that ML readers would look for first, the optical module alone replaced by a standard digital module with everything else unchanged, is rarely among them. Among 16 recent task-level\footnote{Operation-level papers, in which the optics implements a named, mostly linear operation, have also advanced considerably in recent years (e.g., Refs.~\cite{ahmed2025universal,hua2025integrated}), and the questions about system-level claims apply to them too (Sec.~\ref{sec:scope}), but they are not included here.} ONN papers focused on computing, in which the optics takes over part of a digital model in a classical ML pipeline (Appendix~\ref{app:census}), none reports an in-place digital substitution that isolates the contribution of the optical module. If we are lenient and also accept a complete digital model that replaces the entire pipeline, when the pipeline's own digital processing is simple, such as a linear classifier or ridge regression, four do so on at least some of their tasks \cite{yildirim2024nonlinear,wang2024large,liu2026nonlinear,ng2026few}. None of the 16 states its system-level claims, such as speed or energy, in a way that isolates the contribution of the optical module itself.
\section{Motivation}
\label{sec:motivation}

\paragraph{In ML, this ablation study is the default.}
When an ML paper proposes a new module, say a better attention mechanism, it puts the new module where the old one was, keeps the rest of the model, the data, the training and the metric the same, and reports whether task performance improves. This is an ablation study, a routine part of ML papers, and it is what tells readers what the new module adds. ONN modules used in ML pipelines are proposed in exactly this way, as components that take the place of digital computation, so the same ablation study can come with them. Two habits of language, which come naturally to optics, make its absence easy to overlook: calling optical elements ``layers'' and ``neurons'', and counting operations from the physics.

\paragraph{When optical ``layers'' are not ML layers.}
Part of the interpretability gap is terminological: the two communities reuse words for different concepts.
For example, in the diffractive neural network platform \cite{lin2018all}, it is natural to call each diffractive phase mask a ``layer'' and each controllable unit on the masks a ``neuron''.
To an ML reader, ``many layers'' and ``millions of neurons'' suggest compositional depth and high expressivity. However, under the original operation-level setting with amplitude encoding, a cascade of phase masks and free-space propagation implements a \emph{single linear transformation} on the input (a \emph{single layer} in ML terms); the number of physical ``layers'' does not imply the compositional depth that ``layers'' conveys in ML.
This terminology is natural for optics (e.g., phase masks are physically stacked layers), but the difference from ML usage should be clarified when addressing ML audiences.

\paragraph{The mismatch becomes more consequential for system-level claims.}
Physics-native operation counts (e.g., derived from pixels, diffractive features, update rates) may not align with ML notions of parameters, multiply-adds, or FLOPs/TOPS for a digital model.
Just as optical ``layers'' and ``neurons'' are not ML layers and neurons, the tunable elements of an optical module are not parameters in the ML sense, and operations counted from them do not tell how much computation a digital model needs for the same task. These device-native metrics can be useful internally, within the optics community, but cross-platform comparison needs a shared reference grounded in what the module does in the pipeline.

A second mismatch concerns system-level claims, such as speed or energy, stated per operation. Digital processors implement standardized numerical operations at specified precision, independent of the neural network in which they are used, although their actual energy and throughput depend on the workload. Analog optical hardware does not generally offer the same network-independent numerical reliability. Even when numerical fidelity is demonstrated for a particular operation (as in some operation-level papers), this does not establish reliable execution of arbitrary neural networks at the same precision, especially deep and large ones. What ONN papers demonstrate, and usually report, is performance on specific tasks. Their system-level advantages are therefore most meaningfully evaluated \emph{per inference} on those tasks, at matched task performance (Sec.~\ref{sec:systemclaims}).

\paragraph{What we need: a shared, ML-native reference.}
As noted in the Introduction, whatever its internal physics, an optical module in a digital pipeline takes in a real-valued tensor and returns one. The first step is therefore to state this mapping plainly: what goes in, what comes out, and their sizes. Once the boundary of the optical module is stated in this way, its empirical expressivity can be measured as for any other ML component, with an ablation study that isolates its contribution: standard digital modules are put in its place, and their task performance is compared with that of the optical module.
Physical quantities such as operation counts, energy, and speed then enter only the system-level claims, stated against the performance-matched digital module.
\section{Scope}
\label{sec:scope}

\paragraph{The target: systems focused on computing.}
We focus on optical modules that sit in a classical ML pipeline as a mapping from real-valued tensors to real-valued tensors, and above all ONN systems that focus on computing: optical modules proposed as AI or ML accelerators, meant to take over computation that digital hardware does today, with system-level claims, such as speed or energy, over digital counterparts. For these, an ablation study with standard digital modules in the optical module's place is already implied by the claim, and the card asks only that it be reported.

\paragraph{Operation-level modules.}
For an optical module that implements a named operation, such as a matrix multiplication, the operation already tells what the module does, and the ablation study is straightforward (Appendix~\ref{sec:caseA}). Its system-level claims raise the same questions as in Sec.~\ref{sec:systemclaims}, however: operations counted from the optical hardware can far exceed those of the same operation done digitally, which is the count that reflects what the module does. We focus on task-level modules in this paper because for them even what the module does, and hence what to count, is unclear without the ablation study.

\paragraph{Systems focused on sensing: less essential, still informative.}
In ONN systems that focus on sensing, optical processing occurs \emph{before} digitization, meaning no digital input tensor exists before the optics. A digital substitute has to start instead from the scene or signal as recorded by a detector, and the advantages claimed are usually over a conventional sensing system as a whole, such as a camera followed by digital processing, rather than over a digital module in the optics' place (Appendix~\ref{app:census}). The ablation study is therefore less essential there. It can still be informative: since these systems compute as they sense, a comparable digital module gives readers a sense of how much computation the optics performs.

\paragraph{Outside the classical pipeline.}
Some settings fall outside the classical pipeline, for example modules whose inputs or outputs are complex-valued fields, or quantum systems that process quantum states. The approach extends naturally to them, although the ablation study may need adapting, for example when no digital module with the same interface exists; we leave these settings outside the scope of this paper. Here we keep to the minimal requirement, so that the card is easy to adopt and brings the question to a broad audience: it takes the task, metric, and pipeline from each paper, removes nothing from its physics, and adds one ablation study.

\paragraph{Beyond optics: physical analog neural networks.}
We focus on ONNs as currently the most prominent example of physical analog AI acceleration. However, the core issue is not optics-specific. Any physical computing module within an ML pipeline (e.g., neuromorphic substrates \cite{markovic2020physics,kudithipudi2025neuromorphic}, analog accelerators \cite{sebastian2020memory,ambrogio2023analog}, physical neural networks more broadly \cite{momeni2023backpropagation,wright2022deep,momeni2025training}, and even physical reservoirs \cite{tanaka2019recent}\footnote{The simplest case: a fixed physical system followed by a trained linear readout. Reservoir computing papers often report results with and without the physical reservoir, which shows that it contributes; we ask in addition which standard module, put in its place, matches it (Appendix~\ref{app:census-comparisons}). For a small reservoir computing demonstration this may seem excessive; it matters more once a physical system claims system-level advantages over the digital computation it would replace, or is put forward for large-scale AI, as optical systems have been \cite{chen2025all,xu2024large}.}) requires the same kind of evaluation. Consequently, the Expressivity Card serves as a general reporting pattern across physical hardware: define the boundary of the physical module in the ML pipeline and report ablation studies that isolate its contribution, to ground the hardware's computational capacity; the matched digital module then gives a fairer anchor for any system-level claim than counts derived from the hardware itself.
\section{Proposal: define the boundary, then run an ablation study}
\label{sec:boundary}
\label{sec:card}

The proposal follows the logic of the previous sections. The optical module is a mapping from real-valued tensors to real-valued tensors, so its boundary in the pipeline is defined first, and an ablation study then isolates its contribution. Neither step needs new machinery: both are what ML papers already do for a new module. The matched digital module that the ablation study identifies then anchors system-level claims (Sec.~\ref{sec:systemclaims}).

\subsection{Defining the boundary of the optical module}
\label{sec:boundarydef}
\paragraph{The module is a real-to-real mapping.}
The surrounding ML computation interacts with the optical module only through (a) a real-valued control tensor sent to hardware and (b) a real-valued measurement tensor returned to the digital pipeline. Therefore, for ML comparison, the optical module induces a mapping
\begin{equation}
\label{eq:fopt}
 f_{\mathrm{opt}}: \mathcal{X} \subseteq \R^{n} \to \mathcal{Y} \subseteq \R^{m},
\end{equation}
where $\mathcal{X}$ and $\mathcal{Y}$ are the allowed inputs and outputs (e.g., a bounded phase range, nonnegative intensities, sensor resolution, normalization, quantization). In practice, $\R^{n}$ and $\R^{m}$ may represent structured tensors; the key point is that the interface is real-valued.
This definition characterizes the optical module as a real-to-real mapping $f_{\mathrm{opt}}$, without requiring ML readers to understand the underlying physics. It does not claim that physics is irrelevant; rather, it separates the effort of \emph{building} the physical module from the effort of \emph{using and evaluating} it within an ML pipeline.

\paragraph{Where the module begins and ends.}
This is usually straightforward: the optical module is where the signal is optical. Digital data have to be turned into light, for example through a lookup table that sets the voltages driving a modulator, which in turn sets the phase or amplitude of the light, or by setting the power of a laser; that is where the module begins. It ends where the light is detected, typically by intensity detection on a camera or photodiode, or by coherent (homodyne) detection. Appendix~\ref{app:interface} lists common encodings and readouts and the constraints they place on $\mathcal{X}$ and $\mathcal{Y}$.
A common confusion is that parts of the pipeline may be ``close to optics'' (e.g., a digital encoder that produces a phase pattern), but they remain ordinary digital computation and sit outside the optical module.
Some systems are more hybrid, for example with electronics between optical stages, or with part of the encoding done in analog electronics. The line can then be drawn strictly, counting everything electronic as outside the module, or loosely, grouping the hybrid parts into one module; either is fine, as long as the choice is stated. Because the system sits in a digital pipeline, there is always a digital input and output somewhere, and the ablation study can be done there; and an interface designed around the optics, if anything, favors the optical module (Sec.~\ref{sec:fairness}).

\paragraph{Defining the boundary is part of proposing the module.}
When a paper proposes a new module for an ML pipeline, a clearly stated boundary is what lets the module's contribution be isolated, and the authors know their system best. We therefore recommend that ONN papers state the boundary of their optical module plainly, without optics jargon: what goes in, what comes out, and their sizes and ranges. The exact form matters less than having it.

\subsection{Empirical expressivity, by an ablation study}
Once the boundary is stated, the module's empirical expressivity can be measured. This is already how most ONN papers demonstrate their systems: through task performance, for example classification accuracy on MNIST, CIFAR-10, or ImageNet, or a score for generated images. The only addition is the question an ML reader asks next: which standard digital modules, put in the optical module's place, reach the same performance? Answering it is an ablation study that isolates the optical module's contribution, as in Sec.~\ref{sec:motivation}.
\label{sec:substitution}
A simple protocol:

\textit{Step 1: Inherit the task-level pipeline.}
Maintain the exact task, dataset, evaluation metrics, and surrounding digital architecture \emph{already chosen} to evaluate the ONN system. Since authors must define this pipeline to report the performance of their own system, this protocol introduces no new subjective choices at the system level. The goal is simply to evaluate the optical module exactly \emph{as used} within the proposed system, rather than redesigning the overarching model.

\textit{Step 2: State the module's input and output.}
Write down the real-valued input and output tensors, along with any domain constraints, as above. These fix the input and output sizes that all digital substitutes must match.

\textit{Step 3: Select a range of standard digital substitutes.}
Replace the optical module with a set of digital substitutes with the same inputs and outputs.
The substitute architectures should be empirically chosen such that at least one nearly matches the optical module's performance. In addition, the full set should span a meaningful capacity range, including at least one substitute clearly worse than the optical module and one clearly better.
We acknowledge this is an empirical process and recommend including a range of architectures for a comprehensive view; in practice, a ladder of familiar families (linear, MLP, CNN) suffices.
A natural stopping criterion is reached when an expressive substitute drastically outperforms the optical module; then there is no need to search further.
Even a single familiar module, such as a linear layer, a small MLP, or an architecture commonly used for the task, already gives a useful reference: some reference is far better than none.

\textit{Step 4: Train with standard ML practice.}
Train each substitute end-to-end using standard ML practice. Report the training procedure (optimizer, schedule, epochs/steps, key hyperparameters) so that results are reproducible. Standard defaults typically suffice; if extensive tuning is performed, report the protocol.
Several seeds and a wide sweep, as in our case study, are not required; a few single runs already give the picture the card asks for.

\textit{Step 5: Report task performance for all substitutes.}
Report the task performance of each digital substitute using \textit{the same metrics} applied to the optical module. These results ground the optical module's empirical expressivity relative to familiar digital architectures, allowing readers to independently assess which familiar module it compares with.

\paragraph{One way to summarize it: an Expressivity Card.}
For ease of communication, the definition of the module and its empirical expressivity can be summarized in a short card, the Expressivity Card (EC). The box below is one example of its format; what matters is that the information is reported, not the format. The card is deliberately minimal so that it can be easily added to existing ONN works that already have an established architecture and evaluation protocol. Device-level details (fabrication, optical layout, calibration) remain essential for optics audiences but belong in the main technical narrative, not the card.
 The substitutes are matched on task performance, not on parameter count or FLOPs; the matched substitute's parameters and FLOPs are an output of the card (Appendix~\ref{sec:paramcount}). Section~\ref{sec:casestudies} gives a filled-in card.

\begin{expressivitycard}
\small
\begin{enumerate}[leftmargin=1.5em, itemsep=0.2em, label=\textbf{\arabic*.}]
    \item \textbf{Input:} $x \in \mathcal{X}$ (tensor shape; domain constraints if applied).
    \item \textbf{Output:} $y \in \mathcal{Y}$ (tensor shape; domain constraints if applied).
    \item \textbf{Empirical expressivity:} Task performance of standard digital substitutes in the module's place, trained with standard ML practice, using the same evaluation metrics as for the optical module.
\end{enumerate}
\end{expressivitycard}

\subsection{Anchoring system-level claims}
\label{sec:systemclaims}
Most ONN papers are motivated by system-level benefits, such as latency, throughput, and energy efficiency. This is where the card is most directly useful (point~3 in Sec.~\ref{sec:intro}): once the card is in hand, system-level claims can be stated against the digital module that matches the optical module's task performance. Two questions decide what such a claim means: how the operations are counted, and whether the claim is stated per operation or per inference on the demonstrated task.

\paragraph{Counting operations.}
Evaluating an optical module by its task performance needs no operation count, but many system-level claims rest on one, for example a throughput in operations per second or an energy per operation. Operations can be counted in the optical hardware, and such counts are legitimate in their own terms. As Sec.~\ref{sec:motivation} explains, however, they are not equivalent to the operations of the digital module that the optics would replace: optical ``layers'' are not ML layers, and the tunable elements of an optical module are not parameters in the ML sense. How, then, should the operations be counted? Published claims mostly take one of two routes. Some count directly from the parameters of the optical hardware, for example every input pixel connected to every output pixel, or each diffractive layer counted as a full matrix multiplication. Others count the cost of simulating the physics on a digital processor, that is, of the light propagation inside the module of Fig.~\ref{fig:teaser} rather than of the mapping at its interface. Appendix~\ref{app:census} marks these two physics-native counts \lH{} and \lS{}. Neither is a faithful comparison for a module in an ML pipeline, whose contribution is shown by its task performance. Counting is not straightforward, but the card provides an anchor: the matched substitute does the optical module's job in its place, so the operations it needs per inference, times the optical module's inference rate, give an effective throughput that reflects the optical module alone, to be set against the rate at which a digital device runs that substitute.

\paragraph{Per inference on the demonstrated task, not per operation.}
System-level claims, such as speed or energy, are often stated per operation and compared with GPU specifications. While per-operation metrics are useful for characterizing hardware, extending them to other neural networks requires reliable numerical behavior across those networks. This has not, to our knowledge, been demonstrated for analog optical hardware, particularly for deep and large networks (Sec.~\ref{sec:motivation}). The evidence an ONN paper provides is its performance on the demonstrated task, so system-level advantages are best stated per inference on that task, at matched reported task performance. The digital reference is a device running the performance-matched substitute, with its energy and latency measured or estimated under specified hardware assumptions. Appendix~\ref{app:census} marks claims stated per operation \lP{}.

\paragraph{An anchor that isolates the optics.}
If an optical module achieves performance comparable to, say, a two-layer MLP in the same place, then system-level claims, such as energy, throughput, or latency, are most interpretable when made against that specific digital module, rather than against physics-native operation counts or the cost of simulating the optics.
This is what the Expressivity Card provides: a fair reference that anchors system-level claims to the optical module itself, isolating its contribution so that readers from different backgrounds can see it clearly. Stating these claims belongs to the ONN paper itself, whose authors know how the system is deployed; the card supplies the reference. Appendix~\ref{app:census} discusses how recent ONN papers count their claims, and against which digital models.
\section{Case study}
\label{sec:casestudies}
\label{sec:caseB}

We now build a card for an existing system. What follows is the complete card exactly as it would appear in a published ONN paper, and there is \emph{no additional methodology beyond this}.
We select the Optical Generative Model (OGM) reported in \cite{chen2025optical} as a notable, representative example of recent task-level ONN architectures.\footnote{We acknowledge that the authors of Ref.~\cite{chen2025optical} released their code and data, which let us replicate their pipeline and build the card after publication. In general, the authors of a system are best placed to report its card, since they already have the full pipeline at hand (Appendix~\ref{sec:morecases}).}

\begin{figure}[t]
    \centering
    \includegraphics[width=\columnwidth]{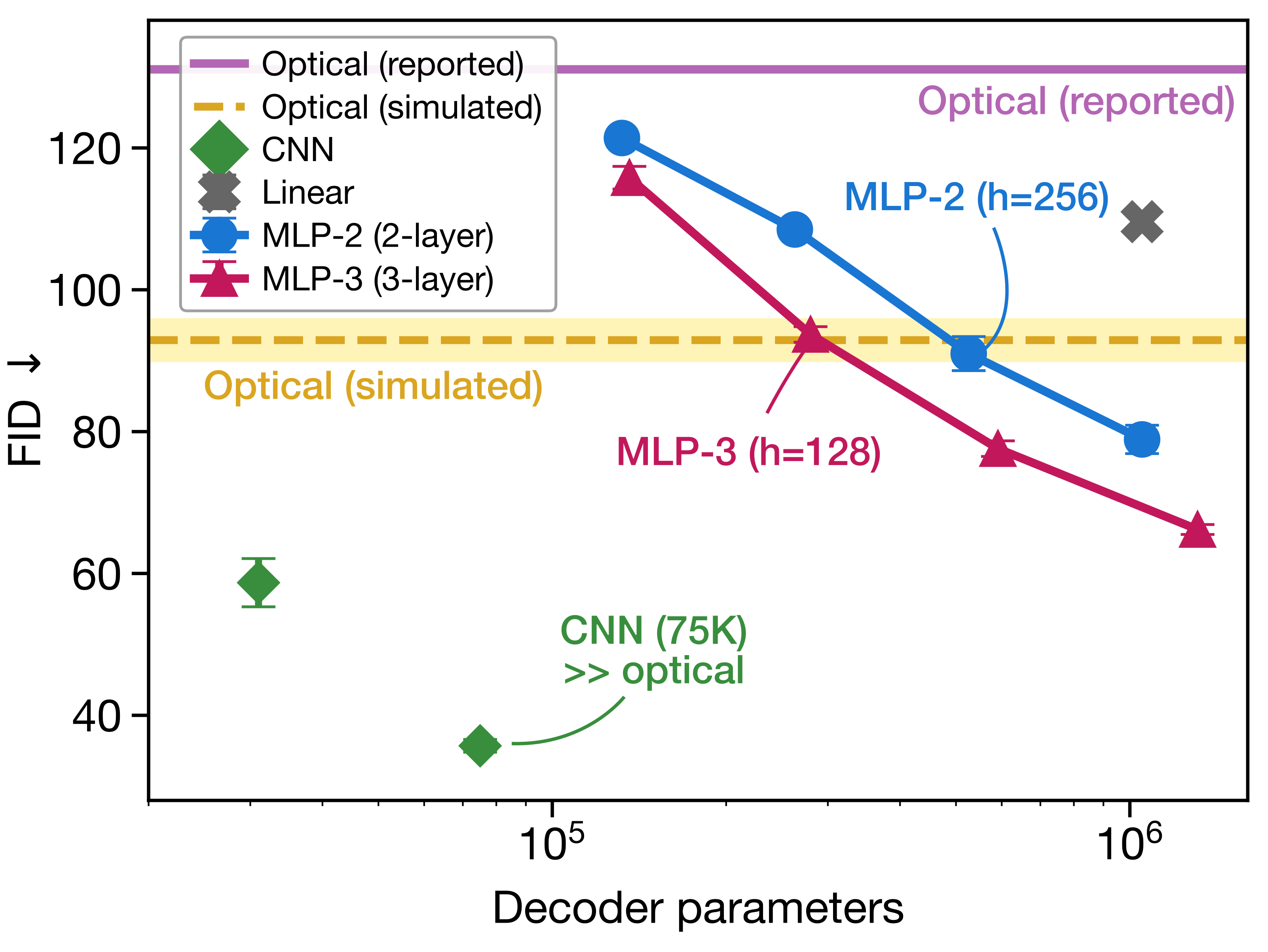}
    \caption{\textbf{Ablation study for OGM, MNIST generation.} FID (lower is better) against the number of trainable parameters of the module in the optical module's place. Markers: digital substitutes, all trained with the same encoder architecture, data, loss, and training setup; mean $\pm$ std over $4$ seeds, FID on $5{,}000$ samples at epoch~$150$. Dashed line and yellow band: the optical module simulated with the authors' released code under the same setup (FID $92.9 \pm 2.9$). Solid line: the experimental FID reported by the authors (131.08), computed on $1{,}000$ generated images and therefore not strictly comparable with the other values. The optical module's $160$K phase-mask pixels are not parameters in the ML sense and are not placed on the horizontal axis.}
    \label{fig:caseB}
\end{figure}

\begin{expressivitycard}{\; OGM (MNIST generation)}
\small
\begin{enumerate}[leftmargin=1.5em, itemsep=0.55em, label=\textbf{\arabic*.}]
    \item \textbf{Input:}~~$\phi \in [0, 2\pi]^{32 \times 32}$ \, (phase pattern, upsampled $10{\times}$ to a $320{\times}320$ SLM in hardware).

    \item \textbf{Output:}~~$y \in \mathbb{R}_+^{32 \times 32}$ \, (intensity image, downsampled $10{\times}$ from a $320{\times}320$ sensor).

    \item \textbf{Empirical expressivity:}~~Optical D$^2$NN (simulated, FID $92.9 \pm 2.9$) falls in the shallow-MLP regime:
    \begin{itemize}[leftmargin=1.5em, itemsep=0.1em, topsep=0.25em, label=$\circ$]
        \item Weaker: Linear (FID $109.6 \pm 1.4$).
        \item Performance-matched: MLP-2 at $h{=}256$ (FID $91.0 \pm 2.4$); likewise, MLP-3 at $h{=}128$ (FID $93.7 \pm 1.1$).
        \item Stronger: $75$K CNN (FID $35.7 \pm 0.9$).
    \end{itemize}
    See Fig.~\ref{fig:caseB} and Table~\ref{tab:caseB} for the full sweep.
\end{enumerate}
\end{expressivitycard}

OGM targets single-shot image generation: once trained, one optical forward pass produces an output image. A digital encoder converts random noise (with optional class conditioning) into a 2D phase pattern displayed on a spatial light modulator (SLM). Light propagates through a diffractive decoding layer (a single trained phase layer, the simplest diffractive deep neural network, D$^2$NN \cite{lin2018all}), and the resulting intensity pattern is captured by a sensor.
The OGM authors also point to uses in which the image need not be digitized at all, such as displays \cite{chen2025optical}; such advantages lie outside what the card measures.

\paragraph{Task and interface.}
We choose the MNIST generation task reported in \cite{chen2025optical}. ML readers know what a generation score on MNIST means. We report the paper's metric, the Fr\'echet Inception Distance (FID; lower is better) \cite{heusel2017gans}.
The digital encoder (a 3-layer class-conditional MLP with 3.3M parameters) outputs a $32 \times 32$ phase pattern. Following Sec.~\ref{sec:boundary}, the optical module maps the phase-control tensor to the measured intensity tensor: the encoder is outside the module, and the SLM, the light propagation, and the camera are inside. For the ablation study, we operate at the $32 \times 32$ interface where the digital encoder and loss computation connect (see Appendix~\ref{app:caseB} for details); the fixed $10\times$ up- and down-sampling around the optics have no trainable parameters.

\paragraph{Why the expressivity is not obvious.}
The OGM paper describes the optical core in physics-native terms (phase modulation, diffraction, intensity detection), as is natural for an optics audience, and motivates phase encoding as a nonlinear encoding that outperforms amplitude or intensity encoding. Because the interface includes phase encoding and magnitude-squared detection, the induced mapping is indeed nonlinear \cite{wanjura2024fully,li2024nonlinear}, but its effective capacity is not obvious from physics diagrams, and estimating it from the physics alone is hard even for experts.
An ML reader cannot readily determine whether to conceptualize the module as a constrained linear map, a shallow MLP, or something qualitatively different.

\paragraph{Results of the ablation study.}
We evaluate digital substitutes at the $32{\times}32$ interface across three families---Linear, MLP, and CNN. For MLPs, we sweep the hidden dimension $h \in \{64, 128, 256, 512\}$ at depths $2$ and $3$ (MLP-2 and MLP-3; Appendix~\ref{app:caseB}); for CNNs, we report two variants spanning $31$K to $75$K parameters. Each substitute is trained from scratch with one standard training setup shared by all substitutes and no per-architecture tuning, with $4$ random seeds per setting (details in Appendix~\ref{app:caseB}).
The original paper \cite{chen2025optical} reported that the physical experimental implementation achieved an FID value of 131.08, computed on $1{,}000$ generated images; we obtained the simulated result with the authors' code, with FID computed on $5{,}000$ samples as for the substitutes.
Figure~\ref{fig:caseB} and Table~\ref{tab:caseB} summarize the results.

\begin{table}[t]
\centering
\caption{\textbf{Empirical expressivity on MNIST} (FID, lower is better). All digital substitutes are trained under a single fixed configuration with $4$ random seeds, evaluated at epoch~$150$ (mean $\pm$ std). \textit{Optical (reported)}: experimental FID reported in~\cite{chen2025optical}, computed on $1{,}000$ generated images (ours: $5{,}000$); \textit{Optical (simulated)}: $4$-seed run of the authors' released simulator. \textit{Mod.}: parameters of the module in the optical module's place; \textit{Total}: total pipeline parameters.}
\label{tab:caseB}
\setlength{\tabcolsep}{4pt}
\resizebox{\columnwidth}{!}{%
\begin{tabular}{lccc}
\toprule
\textbf{Module } & \textbf{Params (Mod.)} & \textbf{Params (Total)} & \textbf{FID} $\downarrow$ \\
\midrule
\rowcolor{gray!15} \multicolumn{4}{l}{\textit{\textbf{Optical module}}}\\
Optical (reported, hardware) & 160K\textsuperscript{*} & 3.3M & 131 \\
Optical (simulated)           & 160K\textsuperscript{*} & 3.3M & $92.9 \pm 2.9$ \\
\midrule
\rowcolor{gray!15} \multicolumn{4}{l}{\textit{Digital Substitutes (Linear)}} \\
Linear & 1.05M & 4.2M & $109.6 \pm 1.4$ \\
\midrule
\rowcolor{gray!15} \multicolumn{4}{l}{\textit{Digital Substitutes (MLP-2)}} \\
MLP-2 ($h{=}64$)   & 132K  & 3.3M & $121.4 \pm 0.6$ \\
MLP-2 ($h{=}128$)  & 263K  & 3.5M & $108.5 \pm 0.9$ \\
MLP-2 ($h{=}256$)  & 526K  & 3.7M & $91.0 \pm 2.4$ \\
MLP-2 ($h{=}512$)  & 1.05M & 4.2M & $78.9 \pm 2.0$ \\
\midrule
\rowcolor{gray!15} \multicolumn{4}{l}{\textit{Digital Substitutes (MLP-3)}} \\
MLP-3 ($h{=}64$)   & 136K  & 3.3M & $115.8 \pm 1.6$ \\
MLP-3 ($h{=}128$)  & 280K  & 3.5M & $93.7 \pm 1.1$ \\
MLP-3 ($h{=}256$)  & 591K  & 3.8M & $77.6 \pm 1.1$ \\
MLP-3 ($h{=}512$)  & 1.31M & 4.5M & $66.2 \pm 0.7$ \\
\midrule
\rowcolor{gray!15} \multicolumn{4}{l}{\textit{Digital Substitutes (CNN)}} \\
CNN ($31$K)   & 31K  & 3.3M & $58.7 \pm 3.4$ \\
CNN ($75$K)   & 75K  & 3.4M & $\mathbf{35.7 \pm 0.9}$ \\
\bottomrule
\multicolumn{4}{l}{\scriptsize \textsuperscript{*} Trainable diffractive features (phase pixels), not standard ML parameters.}
\end{tabular}}
\end{table}

Two patterns stand out. \textit{(i)} The simulated optical system (FID~$92.9 \pm 2.9$) is closely bracketed by an MLP-2 at $h{=}256$ (FID~$91.0 \pm 2.4$) and an MLP-3 at $h{=}128$ (FID~$93.7 \pm 1.1$). This places the optical module's empirical expressivity at the level of \emph{a shallow MLP}.
\textit{(ii)} A $5$-layer CNN with $75$K parameters reaches FID~$35.7 \pm 0.9$, outperforming the optical module by more than $50$ FID points. A small standard CNN therefore performs substantially better in the same place, so the search can stop here (Step~3).

The card reveals that a D$^2$NN \cite{lin2018all} utilizing phase encoding \cite{wanjura2024fully,li2024nonlinear} has an empirical expressivity comparable to a shallow MLP. While this result might not surprise experts within the ONN community, it was not obvious in the original paper. Because the optical module was presented as the decoder of a generative model, trained via knowledge distillation from a large diffusion model (details in Appendix~\ref{app:caseB}), its inherent capacity remained ambiguous without an ablation study in the exact same pipeline.
Furthermore, this ablation study leans toward the optical module (Sec.~\ref{sec:fairness}), so if anything the match flatters it.

More importantly, the matched substitute provides a performance-matched anchor for downstream system-level claims, such as speed or energy.
Concretely, the matched MLP-3 ($h{=}128$) and MLP-2 ($h{=}256$) need about $0.56$ and $1.05$ million floating-point operations (FLOPs) per $32\times32$ image, far fewer than simulating the light would take (Appendix~\ref{app:caseB}). Their energy and latency per inference depend on the digital hardware chosen for deployment, which the ONN authors know best; the card supplies the model to evaluate (Sec.~\ref{sec:systemclaims}).
\section{Alternative Views}
\label{sec:altviews}

We hope to invite discussion and welcome different perspectives. We therefore address several common questions and alternative views about the Expressivity Card below, with a more extended discussion provided in Appendix~\ref{sec:discussion}.

\subsection{ONNs are physical systems; physics-first descriptions should remain primary.}

\textbf{Response.}
We agree that physics-first descriptions are essential for designing, fabricating, and scaling optical systems. We do not claim which is more ``primary''; both serve different audiences and purposes. Physics details matter most to the ONN engineers who build these systems; a general ML audience needs, at the least, what the module does in the pipeline. Physics-first evaluation is essential to most ONN papers and should remain so. Our claim is \emph{not} to replace physics, but to add a \emph{minimal translation layer} ensuring that when an ONN is claimed as an ML accelerator, its contribution to the pipeline is made explicit alongside the physics. Without a description of the interface, ML readers cannot interpret or compare claims, and ONN researchers cannot easily communicate across platforms. Expressivity Cards keep physics discussions intact while making the ML interface explicit.

\textbf{But why do we emphasize the function-oriented evaluation?}
The asymmetric emphasis is a deliberate response to what is \textit{currently underemphasized in the literature}.
Consider a \textbf{GPU analogy}: we evaluate a GPU by what functions it realizes and how efficiently, not by how electrical current flows through the transistors. The same principle applies to optical modules claimed as ML accelerators.

\subsection{Empirical expressivity depends on the specific task, metric, pipeline, etc.; does the card depend too much on these choices?}
\label{sec:altviews_task}

\textbf{Response.}
The EC is, by design, empirically determined by these choices. It does not propose a benchmark; it is a minimal add-on to what the ONN paper \emph{already reports}. The task, metric, and pipeline are chosen by the ONN authors themselves, and the EC simply follows those choices.
A related concern is that the card conflates function class, inductive bias, and trainability. This is an inherent property of any empirical ML evaluation, not a flaw specific to the EC. The EC is transparent about this: it reports empirical performance on the authors' declared task, not a universal expressivity measure.

\subsection{Function-oriented evaluation is standard in ML; what is the contribution of this paper?}

\textbf{Response.}
Function-oriented evaluation, in the form of ablation studies, is indeed the default in the ML community. This is partly because ML researchers often already share a common intuition about the expressivity and limitations of standard architectures, which makes it natural to evaluate new modules by comparing their input--output behavior against established standard architectures.
However, this practice is notably absent in task-level ONN papers, as evidenced by recent high-impact publications: of the 16 papers focused on computing discussed in Sec.~\ref{sec:background}, none reports an ablation in which only the optical module is replaced by a standard digital module, and none states its system-level claims, such as speed or energy, in a way that isolates the contribution of the optical module (Appendix~\ref{app:census}).
The boundary of the optical module is sometimes not clear either, for example whether the electronics that drive the modulators belong to it. This is largely a technical detail, but because the optical module sits in a digital pipeline anyway, stating its boundary plainly costs little and tells a general reader what the module is (Sec.~\ref{sec:boundary}).
Consequently, it is also difficult for ML readers to accurately gauge the actual contributions of the optical modules in ONN papers.

This creates a paradox: ONNs represent a billion-dollar (Appendix~\ref{app:industry}), ML-adjacent field that most mainstream ML researchers cannot meaningfully engage with. Our paper targets exactly this \emph{accessibility barrier}. By advocating for a simple, standardized reporting norm, we provide a reporting template that makes ONN contributions immediately interpretable to ML audiences without requiring a background in physics.
This does not diminish the importance of physics or hardware; rather, it makes ONN contributions more legible to ML audiences by separating the question of \emph{what function is implemented} from the question of \emph{how that function is physically realized}.

\subsection{Is a matched digital substitute equivalent to the optical module? Does the card miss what matters about the optics?}
\label{sec:equivalence}

\textbf{Response.}
No equivalence is claimed. What the card asks for is an ablation study, the kind ML papers routinely report for a new module: standard digital modules are trained in the optical module's place and compared with it on task performance. It does not ask for a digital module that reproduces the optical mapping point by point, nor for a closed-form substitute. Task-level ONN papers do not claim such equivalence either: they propose the optical module to take over computation done digitally today, and show its value through task performance on an existing ML task. The card uses the same measure, chosen by the authors, not by us (Sec.~\ref{sec:altviews_task}).
Expressivity is multi-faceted, and any compact summary risks oversimplification. The Expressivity Card is intentionally modest: it does \emph{not} claim a complete theoretical characterization, and it does not aim to generalize beyond the paper's declared task and pipeline.
Nor does the card take anything out of the paper: the physics, the inductive biases, and the design insights stay where they are. Noise, drift, and hysteresis, as in any physical hardware (a GPU included), are already reflected in the module's measured performance, so the substitutes do not need to model them (Appendix~\ref{subsec:noise}).

\subsection{Isn't training the substitutes too costly, especially as optical modules scale up?}
\label{sec:cost}

\textbf{Response.}
The starting point matters here. So far, ONN papers take existing ML tasks, for which digital solutions are mature: the architectures, training recipes, and data pipelines are standard. Training the substitutes is therefore an ordinary ablation study, with standard ML practice, a short ladder of familiar modules, and a stopping criterion (Sec.~\ref{sec:substitution}). Our case study did more than a card needs, with four seeds per setting for error bars and a wide sweep; in all, it took about 50 GPU-hours (Appendix~\ref{app:caseB}). Scaling up does not change this. Optical modules demonstrated so far are far smaller than state-of-the-art digital models, so the substitutes that match them are typically small by ML standards. We would welcome the day optical systems outgrow state-of-the-art digital models, but that day seems far off. If anything, the effort runs the other way: ONN authors work hard to train their optical modules on these tasks, while training standard digital modules on the same tasks is routine. And if the claim is that an optical module is easier to optimize than its digital counterpart, that is a system-level claim, to be stated as such (Sec.~\ref{sec:sysadv}).

\subsection{Does the card overlook system-level advantages such as energy, latency, and throughput?}
\label{sec:sysadv}

\textbf{Response.}
No; stating these advantages clearly is what the card is for. This is point~3 in Sec.~\ref{sec:intro}: system-level claims, such as speed or energy, are stated per inference against the digital module that matches the optical module's task performance, and the ablation study is what identifies that module (Sec.~\ref{sec:systemclaims}, Appendix~\ref{sec:discussion_resources}). An advantage in energy or latency then compares two modules that do the same job in the same place, a claim any ML reader can check. The card thus makes system-level advantages more clearly stated and more convincing.

\subsection{Is the ablation study fair to the optical module?}
\label{sec:fairness}

\textbf{Response.}
If anything, the EC is more likely to be unfair to a simple digital substitute than to the optical system: the tasks, metrics, boundary, and surrounding digital pipelines in ONN papers are typically chosen by the ONN authors to highlight the optical module's capabilities, and we simply ask what a standard digital substitute achieves under this already-favorable context. Furthermore, ONN authors typically invest substantial effort optimizing their platforms to report the results. We believe a reference biased toward the optical module is still substantially more informative than no reference at all.

A natural concern is that digital substitutes might be trained ``better'' than the physical optical module due to highly mature ML optimization tools. However, because ONN papers fundamentally claim advantages over standard digital counterparts, the digital substitutes must represent what a competent ML practitioner would actually achieve using common digital tools. Therefore, the correct default is to train the digital substitutes exactly as an ML researcher would normally train a model, as in any ablation study. We deliberately stay on the safe side of this default: in our case study, no substitute is tuned for its own architecture, and the ablation study uses the simulated optical result evaluated under the same protocol rather than the experimental one (Sec.~\ref{sec:caseB}, Appendix~\ref{app:caseB}).

Evaluating against a standard, normally trained digital counterpart also provides actionable diagnostic value for optics researchers. If a simple digital module matches the ONN's performance, it sends a clear signal: either the optical module's computational capacity is limited under the current interface, or the physical system suffers from an optimization bottleneck during training. Both are crucial insights for progressing the hardware design.

\section{Conclusion}
\label{sec:conclusion}

ONN papers often contain significant systems insights about unconventional computational platforms, and our intent is not to diminish the engineering contributions of the ONN community. Quite the opposite: we aim to make these efforts more accessible to the ML community. Without a shared description of what the module does, readers can disagree (or talk past each other) about what a given result implies. A function-oriented norm benefits:
\begin{itemize}
\item \textbf{ML researchers,} by turning an optical subsystem into a well-defined building block that can be compared, substituted, and integrated into standard ML workflows without requiring a deep physics background.
\item \textbf{ONN researchers,} by providing a common language for comparing designs, identifying which tasks best match a module's capabilities, and tracking whether new work increases expressivity rather than merely changing the demonstration task or accounting convention.
\end{itemize}

Optical hardware for ML is advancing rapidly, both in operation-level accelerators and in task-level optical modules embedded in more sophisticated pipelines. As these systems become more capable, clear evaluation norms become increasingly important. We propose that optical modules used within classical ML pipelines should be evaluated and communicated by their ML-relevant function, not only by their physics. To make their contribution clear, we should define the boundary of the optical module in the pipeline and report an ablation study that puts standard digital modules in its place, as is routine for new modules in ML. This allows readers to understand what the optics contributes without needing to know the physics, and provides a performance-matched digital reference for system-level claims, such as speed or energy, stated per inference on the demonstrated task. 
The Expressivity Card offers one simple way to add this evaluation with little extra effort. We do not need an exhaustive study to make progress, and even a simple comparison with one digital module in the optics' place is a useful first step.

\clearpage
\bibliography{references}
\bibliographystyle{unsrtnat}

\clearpage
\appendix
\onecolumn
\section{Reporting practice in recent task-level ONN papers}
\label{app:census}
\newcommand{\vyes}{\textcolor{verdictyes}{Yes}}
\newcommand{\vyesp}{\textcolor{verdictyes}{Yes (partly)}}
\newcommand{\vweak}{\textcolor{verdictweak}{Weaker yes}}
\newcommand{\vweakp}{\textcolor{verdictweak}{Weaker yes (partly)}}
\newcommand{\vno}{\textcolor{verdictno}{No}}
\newcolumntype{L}[1]{>{\raggedright\arraybackslash}p{#1}}
\newcommand{\inplace}{\textbf{\emph{in place}}}
\newcommand{\Inplace}{\textbf{\emph{In place}}}
\newcommand{\rowsep}{\addlinespace[1.4pt]\arrayrulecolor{black!20}\midrule[0.3pt]\arrayrulecolor{black}\addlinespace[1.4pt]}

We chose recent task-level ONN papers that we read closely and consider representative\footnote{We did not include operation-level papers (Sec.~\ref{sec:background}): for them the operation itself tells what to count, although their system-level claims are not always counted from it (Sec.~\ref{sec:scope}). Some people use ``ONN'' more strictly, only for systems whose optical modulation is optimized for the task. We use a broader definition, which also includes systems whose optics is fixed (e.g., \cite{xia2024nonlinear,wang2024large}), because the line between the two is not always sharp and the question we ask is the same for both.} (Sec.~\ref{sec:background}) and grouped them by focus as in Sec.~\ref{sec:scope}: 16 focused on computing (Table~\ref{tab:census-full}) and 15 focused on sensing (Table~\ref{tab:census-sensing}). This is not a complete survey: we may have missed related papers, and we would welcome suggestions. Papers that propose methods for training or designing optical modules \cite{xue2024fully,wanjura2024fully,li2024nonlinear} advance how such modules are built; we do not include them because they propose methods rather than systems. We consider all of the included papers good work, and if their authors see any point differently, we would be glad to hear from them and learn more.

For each paper, the tables list its main tasks, its system-level claims, the digital models it evaluates on its tasks (its digital baselines), and whether these isolate the contribution of the optical module. The entries use short labels, defined in Appendices~\ref{app:census-claims}--\ref{app:census-isolated}:
\begin{itemize}[leftmargin=1.5em, itemsep=0.15em]
\item Each system-level claim carries a letter for where its operation count comes from: \lH, the optical hardware; \lS, simulating the optics; \lD, a digital model for the task; \lF, the system's digital part only, with the optics counted as free; or \lN, no digital counterpart. A second letter, \lP, marks a claim stated per operation rather than per inference.
\item Each digital baseline carries a label for where the digital model sits in the pipeline: \inplace, \emph{whole}, \emph{removed}, or \emph{sim.}
\item The last column says whether the paper isolates the optical module: \vyes, \vweak, or \vno.
\end{itemize}
Appendix~\ref{app:census-summary} summarizes the findings.

\subsection{System-level claims and how they are counted}
\label{app:census-claims}
The third column lists each paper's system-level claims, such as speed or energy. Claims stated only against other optical systems, and general statements about ONNs, are not listed; claims without numbers are marked qualitative. As in Sec.~\ref{sec:systemclaims}, each claim is labeled by two choices: where its operation count comes from, and whether it is stated per operation or per inference.

\paragraph{Where the operation count comes from.}
Operations can be counted in the optical hardware, and such counts are legitimate in their own terms, but they are not equivalent to the operations of the digital module that the optics would replace (Sec.~\ref{sec:motivation}). The first label says where a claim's count comes from:
\begin{itemize}[leftmargin=1.5em, itemsep=0.15em]
\item \lH, the optical hardware: operations counted from the parameters of the optical hardware, its pixels, ``neurons'', matrix elements, or layers; for example, every input pixel counted as connected to every output pixel, or each diffractive layer counted as a full matrix multiplication. Optical parameters are not parameters in the ML sense (Sec.~\ref{sec:motivation}), so such a count is not that of a digital module doing the same job.
\item \lS, simulating the optics: operations counted from the cost of simulating the physics, such as the propagation of light, on a digital processor. Such a count measures how hard the physics is to reproduce, not what the module does in the pipeline (Fig.~\ref{fig:teaser}, Sec.~\ref{sec:systemclaims}).
\item \lD, a digital model: operations, time, or energy counted from a digital model for the demonstrated task, or from a conventional system using one. This need not match the optical system's task performance or isolate the optical module: replacing the whole pipeline also replaces its surrounding digital processing. Isolating the optics requires a performance-matched digital module \inplace{} (Appendices~\ref{app:census-comparisons}, \ref{app:census-isolated}  and \ref{app:census-summary}).
\item \lF, optics counted as free: operations, time, or energy counted for the system's digital part only, with the optics, and the conversions between the optical and digital domains, counted as free. Like taking the optics out of the count, this shows how much digital computation the system needs with the optics in place, but it does not quantify what the optics contributes or costs.
\item \lN, no digital counterpart: a figure for the optical system stated on its own, such as the time light takes to pass through it.
\end{itemize}

\paragraph{Per operation or per inference.}
Digital processors provide standardized numerical operations at specified precision across different networks, although their energy and throughput depend on the workload. Analog optical hardware has not, to our knowledge, demonstrated comparable numerical reliability across arbitrary networks, especially deep and large ones (Sec.~\ref{sec:motivation}). Its system-level advantages are therefore best assessed per inference on the demonstrated task, at matched task performance. The second label marks claims stated per operation:
\begin{itemize}[leftmargin=1.5em, itemsep=0.15em]
\item \lP, per operation: a claim stated in operations rather than per inference, such as a throughput in operations per second or an energy efficiency in operations per joule, usually against the same figure for a digital processor, such as a GPU. Where a paper compares one optical operation, such as a matrix--vector multiplication, with the same operation done digitally, the claim is an operation-level comparison and is marked \lP{} alone. Claims without \lP{} are per inference.
\end{itemize}

\subsection{Digital baselines and what each shows}
\label{app:census-comparisons}
The fourth column lists the digital models that each paper evaluates on its tasks, with a label for where each sits in the pipeline. Each kind answers a real question, but only the first isolates the optical module, and the second does so in a weaker form for systems whose digital processing is simple (Appendix~\ref{app:census-isolated}):
\begin{itemize}[leftmargin=1.5em, itemsep=0.15em]
\item \inplace: a standard digital module, such as an MLP or a CNN, in exactly the optical module's place, with everything else unchanged: the same input, the same digital processing before and after the optics, the same data, and the same metric. This is the ablation study of Sec.~\ref{sec:background} and Fig.~\ref{fig:teaser}. It asks what the optics does: which standard module, put in its place, matches its task performance.
\item \emph{whole}: a complete digital model, or a conventional system, that replaces the whole system, the optics together with the digital processing around it. It asks whether a digital system can do the whole task as well, a fair question about the system; but where the system has substantial digital processing, the answer mixes the optical module's contribution with that of this processing (Sec.~\ref{sec:systemclaims}).
\item \emph{removed}: the optics is bypassed and the remaining digital processing is trained or evaluated directly on the input, or a linear model takes the optics' place, which amounts to the same. Reservoir computing papers, for example, often report results with and without the physical reservoir.
This tests whether the optics helps, but it does not establish the exact contribution of the optical module.
\item \emph{sim.}: a digital simulation of the optics. It shows how faithfully the physics is modeled; its cost reflects the physics being reproduced (\lS{} in Appendix~\ref{app:census-claims}), not what the module does in the pipeline.
\end{itemize}
Unless a tag says otherwise, the conditions of a baseline match those of the optical system: the same task input, training data, and metric. A tag in parentheses says what differs: \emph{other input}, \emph{other training data} (for example, pretraining on a larger dataset), or \emph{cited} (results taken from another paper). We do not apply these conditions strictly: the digital model's own input processing, such as resampling the images, counts as simple digital processing, and matching the training effort is good to see but not required. Optical baselines, such as a random instead of a trained mask, are not listed: they are ablation studies of the optics' own design choices (Appendix~\ref{sec:optbaseline}).

\subsection{Whether the optics is isolated}
\label{app:census-isolated}
The last column, \emph{optics isolated}, says whether a paper reports the ablation study of Sec.~\ref{sec:background}, in terms of the labels of Appendix~\ref{app:census-comparisons}:
\begin{itemize}[leftmargin=1.5em, itemsep=0.15em]
\item \vyes: an \inplace{} baseline on all of the paper's tasks, that is, a standard digital module in exactly the optical module's place, with everything else unchanged.
\item \vweak: a \emph{whole} baseline with matched conditions, when the original system's digital processing is simple---no more expressive than a fully connected layer, such as pooling followed by a linear classifier, or ridge regression. We accept this, leniently, as a weaker form of the in-place ablation study of the optical module.  Optical systems with more expressive digital processing in the pipeline, such as an MLP or CNN, do not qualify.
\item ``(partly)'': added when such a baseline covers only some of the tasks.
\item \vno: none of these.
\end{itemize}

\begin{table*}[p]
\caption{\textbf{Task-level ONN papers focused on computing.} Labels as defined in Appendices~\ref{app:census-claims}--\ref{app:census-isolated}; --: none.}
\label{tab:census-full}

\fontsize{7pt}{7.8pt}\selectfont
\setlength{\tabcolsep}{3pt}
\begin{tabular}{@{}L{0.088\textwidth}L{0.163\textwidth}L{0.335\textwidth}L{0.306\textwidth}L{0.058\textwidth}@{}}
\toprule
Paper & Tasks & System-level claims & Digital baselines on the tasks & Optics isolated \\
\midrule
Xu et al.\ 2024, Science \cite{xu2024large} & Image classification (CIFAR-10, Omniglot, etc.); generation of music and stylized images & \emph{Energy and area efficiency} (\lH,~\lP): diffractive propagation counted as dense matrices between adjacent layers; vs.\ an NVIDIA H100 GPU and other optical systems. \emph{Latency} (\lN): inference time per image. & \emph{Whole} (cited): VGG-16 and other electronic networks, by their reported accuracy on CIFAR-10. & \vno \\ \rowsep
Chen et al.\ 2025, Nature \cite{chen2025optical} & Image generation (MNIST, CelebA, Van Gogh-style art, etc.) & \emph{Operations} (\lF): the digital encoder only, vs.\ all-digital generators. \emph{Operations} (\lS): the decoder equated to the cost of simulating it (reply to reviewers). \emph{Energy} (\lD): per image, vs.\ diffusion models. \emph{Latency} (\lN): light transit through the decoder. & \emph{Whole}: fully connected generators; diffusion models with as many parameters as the encoder; the teacher diffusion model. System's digital part: a neural-network encoder. & \vno \\ \rowsep
Chen et al.\ 2025, Science \cite{chen2025all} & Semantic image generation, style transfer, denoising, etc. & \emph{Throughput, energy efficiency} (\lH,~\lP): linear diffractive layers counted as one dense matrix from all input to all output pixels; vs.\ A100 specifications. \emph{Time, energy per frame} (\lD): measured, vs.\ StyleGAN on an A100 at comparable quality. & \emph{Whole} (other input): separately trained digital models, such as Stable Diffusion, StyleGAN, VGG-19 style transfer, and NeRF. & \vno \\ \rowsep
Xia et al.\ 2024, Nat.\ Photon.\ \cite{xia2024nonlinear} & Image classification and reconstruction; keypoint and pedestrian detection & \emph{Digital operations and parameters} (\lF): the system's digital decoder only, vs.\ CNNs on the same tasks. \emph{Response time} (\lN): per video frame, compared with human reaction time. & \emph{Whole}: CNNs on keypoint and pedestrian detection. \emph{Removed}: linear regression with the same number of modes. \emph{Sim.}: MLPs trained to approximate the optical mapping. System's digital part: an MLP or a linear layer. & \vno \\ \rowsep
Yildirim et al.\ 2024, Nat.\ Photon.\ \cite{yildirim2024nonlinear} & Image classification (MNIST, Fashion-MNIST, Imagenette) & -- & \emph{Whole}: LeNet-5 and a lighter CNN, with comparable accuracy on all three tasks. System's digital part: pooling and a linear layer. & \vweak \\ \rowsep
Wang et al.\ 2024, Nat.\ Comput.\ Sci.\ \cite{wang2024large} & Classification of images (sign language, CIFAR-10, etc.) and graphs; regression & \emph{Throughput, energy efficiency} (\lS,~\lP): operations needed to emulate the fixed optical matrix digitally; vs.\ V100 GPU specifications. \emph{Digital operations} (\lF): the linear readout only, vs.\ LeNet-5 at similar accuracy. & \emph{Whole}: LeNet-5 (images); classical regressors (not neural networks). \emph{Whole} (cited): graph models of other papers. \emph{Removed}: a simulated linear random projection. \emph{Sim.}: a simplified model of the optics. System's digital part: a linear readout. & \vweakp \\ \rowsep
Zhan et al.\ 2024, Nat.\ Commun.\ \cite{zhan2024photonic} & Image generation (MNIST) & \emph{Digital operations} (\lF): those of the system's digital part only, vs.\ an MLP of comparable performance (simulation). & \emph{Whole}: an MLP of comparable performance, found by architecture search (simulation). System's digital part: a shallow MLP. & \vno \\ \rowsep
Liu et al.\ 2026, Sci.\ Adv.\ \cite{liu2026nonlinear} & Classification (MNIST, Chinese MNIST, EMNIST letters); image XOR & \emph{Energy} (\lF): the digital readout only, estimated for an embedded GPU; vs.\ the energy reported for digital nonlinear networks in other papers. & \emph{Whole}: a fully connected ReLU network, on all tasks. \emph{Removed}: a linear network. System's digital part: a linear readout. & \vweak \\ \rowsep
Ng et al.\ 2026, Sci.\ Adv.\ \cite{ng2026few} & Few-shot image classification (MNIST, BreaKHis, etc.); skin-lesion segmentation & \emph{Energy} (\lD): the whole system per frame, vs.\ a GPU running a large CNN. \emph{Training time} (\lD): vs.\ EfficientNetV2-B0 at matched accuracy. \emph{Training data} (\lD): more accurate than the digital models with few training images. & \emph{Whole}: an MLP, LeNet-5, and CNNs; pretrained EfficientNetV2-B0 and ViT (other training data). \emph{Removed}: logistic or ridge regression on raw images. System's digital part: logistic or ridge regression. & \vweak \\ \rowsep
Fu et al.\ 2023, Nat.\ Commun.\ \cite{fu2023photonic} & Iris classification; MNIST, compressed to 10 features by a digital layer & \emph{Throughput, energy per operation} (\lH,~\lP): each diffractive layer counted as a full matrix multiplication at an assumed detection rate; vs.\ GPUs, a TPU, and other processors. \emph{Latency} (\lN): light transit through the chip. & -- & \vno \\ \rowsep
Wang et al.\ 2024, Nat.\ Commun.\ \cite{wang2024ultrafast} & Time-series prediction, system emulation, X-ray image classification & \emph{Throughput, energy efficiency} (\lH,~\lP): counted from the complex matrix operations of the star coupler and the weight banks; vs.\ an NVIDIA H100 GPU. & \emph{Sim.}: digital simulations of the optical system. & \vno \\ \rowsep
Aadhi et al.\ 2025, Nat.\ Commun.\ \cite{aadhi2025scalable} & Delayed XOR, time-series prediction, telecom signal recovery, run in parallel & \emph{Throughput, energy per operation} (\lH,~\lP): multiply--accumulates counted from the physical reservoir's nodes and input encoding; vs.\ DSP and electronic networks reported in other papers. & \emph{Removed}: the received signal without processing. \emph{Whole} (cited): DSP and electronic networks from other papers. & \vno \\ \rowsep
Parto et al.\ 2026, Sci.\ Adv.\ \cite{parto2026high} & Time-series prediction, channel equalization, waveform classification & \emph{Latency} (\lN,~\lP): sub-nanosecond per inference if implemented all-optically (projected); vs.\ one clock cycle of a digital processor. & \emph{Removed}: direct detection of the distorted signal; a linear equalizer applied to the input (equalization task). & \vno \\ \rowsep
Wang et al.\ 2025, Light Sci.\ Appl.\ \cite{wang2025optical} & Forecasting (Lorenz63, Kuramoto--Sivashinsky); an observer task & \emph{Throughput, energy efficiency} (\lH,~\lP): counted from the optical random projection; stated as not yet competitive with an NVIDIA V100 GPU. & Cubic spline interpolation, on the observer task only (not a neural network). \emph{Sim.}: digital simulations of the optical system. & \vno \\ \rowsep
Petrini et al.\ 2026, Nature \cite{petrini2026degree} & Color-image classification (CIFAR-10) & \emph{Speed}, qualitative: images classified fully in parallel; the modulator can be integrated into low-latency networks. & -- & \vno \\ \rowsep
Gao et al.\ 2023, Nat.\ Commun.\ \cite{gao2023optofluidic} & XOR; classification of MNIST digit pairs & -- & \emph{Removed}: linear regression on the input, without the physical reservoir. & \vno \\
\bottomrule
\end{tabular}
\end{table*}

\begin{table*}[p]
\caption{\textbf{Task-level ONN papers focused on sensing.} Labels as defined in Appendices~\ref{app:census-claims}--\ref{app:census-isolated}; --: none. In these papers the optics processes light before it is detected, so, unlike in an ML pipeline, which starts and ends with digital data, there is no digital input before the optics (Sec.~\ref{sec:scope}).}
\label{tab:census-sensing}

\fontsize{7pt}{7.8pt}\selectfont
\setlength{\tabcolsep}{3pt}
\begin{tabular}{@{}L{0.088\textwidth}L{0.163\textwidth}L{0.335\textwidth}L{0.306\textwidth}L{0.058\textwidth}@{}}
\toprule
Paper & Tasks & System-level claims & Digital baselines on the tasks & Optics isolated \\
\midrule
Chen et al.\ 2023, Nature \cite{chen2023all} & Image classification (MNIST, Fashion-MNIST, KMNIST, ImageNet classes); traffic video & \emph{Throughput, energy efficiency} (\lH,~\lP): the diffractive masks counted as dense matrices; vs.\ GPU specifications. \emph{The same metrics} (\lD,~\lP): counted as the operations of a LeNet of the same accuracy. \emph{Energy and latency per frame} (\lD): measured, vs.\ fully connected networks and LeNets on an A100 GPU. & \emph{Whole}: fully connected networks and LeNets on the same images (not on KMNIST). \emph{Removed}: the analog electronic layer alone. System's electronic part: a binary-weighted fully connected layer. & \vweakp \\ \rowsep
Wang et al.\ 2023, Nat.\ Photon.\ \cite{wang2023image} & Image classification at the sensor (QuickDraw, cell organelles, etc.) & \emph{Power} (\lD): the whole sensor, estimated at a projected high frame rate; vs.\ a high-speed camera with an FPGA running a digital network. & \Inplace: a digital nonlinear encoder with the same 4-dimensional bottleneck and decoder (QuickDraw); a pretrained ResNet with the same decoder (other training data). \emph{Removed}: a digital linear encoder; direct imaging, downsampled. & \vyesp \\ \rowsep
Peng et al.\ 2026, Nature \cite{peng2026optical} & Object detection, segmentation, 3D reconstruction, video, in natural scenes & \emph{Parameters, operations, energy} (\lF): those of the digital back end only, vs.\ state-of-the-art digital models, some more accurate. \emph{Energy and latency} (\lD): end to end, vs.\ a conventional camera with a digital model. & \emph{Whole}: lightweight and state-of-the-art digital vision models, some pretrained on external data (other training data). System's digital part: a neural-network back end. & \vno \\ \rowsep
Huang et al.\ 2024, Sci.\ Adv.\ \cite{huang2024presensor} & Classification of hand-drawn figures, human actions, and blood cells & \emph{Operations} (\lH): the share done by the optics, counted as those of the convolutions it implements. \emph{Latency} (\lF): the digital post-processing only, vs.\ a digital network of the same structure and similar performance. & -- & \vno \\ \rowsep
Zhou et al.\ 2023, Sci.\ Adv.\ \cite{zhou2023ultrafast} & Spike-sequence classification, action recognition, motion tracking; nanosecond flashing patterns & \emph{Latency} (\lS): vs.\ the same network run on a CPU. \emph{Latency} (\lP), an operation-level comparison: vs.\ one matrix--vector multiplication of similar size on a CPU and a GPU, and the cycle times of FPGAs and event cameras. & -- & \vno \\ \rowsep
Yan et al.\ 2024, Sci.\ Adv.\ \cite{yan2024nanowatt} & Depth perception; obstacle avoidance by a mobile robot & \emph{Power} (\lD): measured per meta-pixel, vs.\ that of stereo vision and lidar. \emph{Latency}, qualitative: depth read out at the speed of light, without the post-processing of stereo vision and lidar. & \Inplace: a U-Net trained on the camera's diffraction images (one task). \emph{Whole}: commercial 3D sensors (a stereo camera, a Kinect, a lidar) on obstacle avoidance. & \vyesp \\ \rowsep
Chen et al.\ 2023, Sci.\ Adv.\ \cite{chen2023photonic} & Image transmission with compression (MNIST, adrenal CT, etc.), encryption, and error correction & \emph{Latency and throughput} (\lD): vs.\ digital DCT compression on electronic processors. & DCT and FFT compression; an error-correcting code (not neural networks). \emph{Removed}: transmission without the encoder--decoder. & \vno \\ \rowsep
Bai et al.\ 2023, Light Sci.\ Appl.\ \cite{bai2023alloptical} & Digit classification through unknown diffusers, single-pixel (terahertz) & \emph{Speed and energy}, qualitative: classification with only the illumination light and no digital computer, unlike digital deep-learning methods. & -- & \vno \\ \rowsep
Huang et al.\ 2026, Light Sci.\ Appl.\ \cite{huang2026anti} & Recognizing a target digit among moving interfering objects (terahertz) & -- & -- & \vno \\ \rowsep
Feng et al.\ 2026, Light Sci.\ Appl.\ \cite{feng2026spatiotemporal} & Classification (MNIST, Fashion-MNIST); lidar detection of moving targets & \emph{Throughput, energy efficiency} (\lH,~\lP): modulation units times the frame rate, times $2^{8}$ for 8-bit data (theoretical); vs.\ the specifications of electronic processors. & -- & \vno \\ \rowsep
Li et al.\ 2023, Nat.\ Commun.\ \cite{li2023rapid} & Detection of hidden defects (terahertz) & \emph{Throughput}, qualitative: inspection without forming, storing, or digitally processing an image, unlike machine-vision pipelines. & -- & \vno \\ \rowsep
Bezuidenhout et al.\ 2026, Nat.\ Commun.\ \cite{bezuidenhout2026deep} & Detecting the skyrmion number of vector beams; sorting OAM modes & \emph{Energy}, qualitative: once trained, the optics could be made as metasurfaces that consume no energy; the authors note that a complete analysis should also include the setup's electrical power. & -- & \vno \\ \rowsep
Tao et al.\ 2026, Light Sci.\ Appl.\ \cite{tao2026nanosecond} & Strain, torsion, deformation, and joint-angle sensing & \emph{Latency} (\lD): vs.\ conventional fiber-sensing systems. \emph{Throughput, energy efficiency} (\lH,~\lP): every SLM pixel counted as connected to every detector pixel. & \Inplace: a U-Net, on the deformation task only. \emph{Sim.}: a simulated diffractive network of the same architecture. & \vyesp \\ \rowsep
Xiao et al.\ 2026, Sci.\ Adv.\ \cite{xiao2026endtoend} & Classification of liquid mixtures; concentration prediction & \emph{Energy and latency} (\lD): estimated per inference, vs.\ a conventional system that stores the full spectrum and then processes it electronically. \emph{Throughput} (\lH,~\lP): counted from the weight bank at the detection rate. & \emph{Whole}: a CNN, on the classification task only. System's inference: in the optical domain. & \vweakp \\ \rowsep
Gu et al.\ 2025, Sci.\ Adv.\ \cite{gu2025allintegrated} & Concurrent sensing of intensity, polarization, and wavelength; recovery of high-speed signals & \emph{Throughput, energy per operation} (\lH,~\lP): counted from the interferometer meshes at the detection bandwidth; vs.\ a digital network of the same size on a GPU. \emph{Processing time, training energy} (\lH): vs.\ that network. & -- & \vno \\
\bottomrule
\end{tabular}
\end{table*}

A baseline counts whether it performs better or worse than the optics; even one familiar digital module is informative. This column concerns ablations on the demonstrated tasks, while Appendix~\ref{app:census-summary} examines whether system-level claims use a performance-matched digital reference.

\subsection{Summary}
\label{app:census-summary}
In terms of these labels, this paper asks for one combination (Fig.~\ref{fig:teaser}, Sec.~\ref{sec:boundary}): a digital baseline \inplace, the standard digital module that, put in the optical module's place, matches its task performance; and system-level claims counted from that same module (\lD), per inference on the demonstrated task. Counts from the optical hardware (\lH) or from simulating the optics (\lS) describe the physics rather than what the module does in the pipeline; counting the optics as free (\lF) leaves out what the optics contributes and costs; and per-operation figures (\lP) do not by themselves establish system-level advantages on ML tasks. 
Tables~\ref{tab:census-full} and~\ref{tab:census-sensing} give the paper-by-paper details; we summarize the main patterns below.

None of the 16 papers focused on computing reports a digital module \inplace{}. Four report a matched \emph{whole} baseline for a system with simple digital processing on some tasks. Among the 15 focused on sensing, three report a digital module \inplace{} on some tasks, and two a matched \emph{whole} baseline for a system with simple digital processing on some tasks.

Throughput and energy-efficiency claims are mostly stated per operation, with operations counted from the optical hardware or from simulating the optics: of the 31 papers (with overlapping categories), 11 count them from the optical hardware (\lH, \lP) and one from simulating the optics (\lS, \lP), and one more makes an operation-level comparison (\lP). Ten state claims counted from a digital model (\lD), mostly per inference, one of them also per operation (\lD, \lP). Seven state claims counted for the digital part of the system only, with the optics counted as free (\lF). A few claims, mostly latencies, have no digital counterpart (\lN). Apart from throughput, energy, and latency, which are the most common, other system-level claims appear too, such as training time and the amount of training data needed.

Few of the papers that isolate the optics also state their system-level claims against the module that matches it. Among the four computing-focused papers with at least a weaker form of isolation, one compares the operations of its digital readout with those of the digital model it compares on the task, with the optics counted as free (\lF); one states no system-level claims; and two state their energy against digital models or figures other than those they compare on the task. Of the five focused on sensing, one states energy and latency per frame against the digital models it compares on the task, at the same accuracy; the other four compare energy, power, or latency with conventional systems instead. None of the 16 papers focused on computing, therefore, states its system-level claims, such as throughput or energy, in a way that isolates the contribution of the optical module itself.

These records describe only what each paper reports, so that the statement in Sec.~\ref{sec:background} can be checked. A No says nothing about the quality of a work: its other baselines answer other, valid questions (Appendix~\ref{sec:discussion}), and many of these papers are, in our view, essential steps in the right direction.

\section{Interface constraints}
\label{app:interface}

A function-oriented report should not require ML readers to learn device physics. However, it should make \emph{interface-induced constraints} explicit, as these constraints determine what functions are feasible.
We recommend describing the interface in two parts: \emph{input encoding} and \emph{output readout}. The specifics of how the optical transformation is physically realized can remain in the paper's technical sections without affecting the evaluation of what the module does.

\subsection{Input encoding: real-valued tensor \texorpdfstring{$\to$}{to} physical control}

Two common input-encoding patterns are:
\begin{itemize}
\item \textbf{Spatial modulation (free space).} A real tensor is mapped to per-pixel electrical controls (typically voltages) on a spatial modulator. The resulting control domain is often phase-only, amplitude-only, or coupled amplitude--phase modulation. The ML-relevant constraints are the allowed range (e.g., phase modulo $2\pi$ or a bounded phase range), quantization/bit depth, and spatial resolution.
\item \textbf{Multi-channel / multi-mode control (integrated photonics).} A real vector is mapped to amplitudes/phases across a set of input channels (waveguides, modes, wavelengths). The ML-relevant constraints include dynamic range, normalization conventions, channel count, and whether the control is signed or restricted (e.g., nonnegative amplitudes).
\end{itemize}
These constraints define $\mathcal{X}$ in Eq.~\eqref{eq:fopt}.

\subsection{Output readout: optical state \texorpdfstring{$\to$}{to} real-valued tensor}

Similarly, the output measurement defines what the digital pipeline receives:
\begin{itemize}
\item \textbf{Direct detection (intensity).} The measurement is nonnegative and often proportional to intensity integrated over pixels/bins. This introduces a magnitude-squared nonlinearity at the interface and yields an output-domain constraint $y\ge 0$.
\item \textbf{Coherent detection (field quadratures).} The measurement returns (approximately) signed real-valued quadratures (the real and imaginary parts of the field) relative to a local oscillator (a reference beam), enabling direct access to phase information at the interface.
\end{itemize}
These define $\mathcal{Y}$ in Eq.~\eqref{eq:fopt} and shape the relevant hypothesis class for $f_{\mathrm{opt}}$.

\subsection{Inside the boundary: the physical core}

The physical core can be realized by a wide range of optical platforms (diffractive layers and metasurfaces, interferometer meshes, microrings, programmable waveguides, etc.), and these choices matter greatly for engineering and scaling. However, the purpose of the Expressivity Card is to ensure that ML readers can interpret the module even if they treat the physical core as a black box. At the interface level, what matters is the realized mapping $f_{\mathrm{opt}}$ and the allowed inputs and outputs $(\mathcal{X},\mathcal{Y})$.
\section{OGM: Full Experimental Details}
\label{app:caseB}

This appendix provides complete experimental details for the Optical Generative Model (OGM) case study in Section~\ref{sec:caseB}.

\subsection{Task Description}

We evaluate the expressivity of the optical diffractive module by comparing it with standard digital architectures on the MNIST dataset~\cite{lecun1998gradient}. Following the OGM framework~\cite{chen2025optical}, the pipeline comprises a digital encoder and a variable decoder module, trained via knowledge distillation from a pretrained Denoising Diffusion Probabilistic Model (DDPM)~\cite{ho2020denoising}.

\paragraph{Teacher model.}
The teacher DDPM is trained with $T=1000$ diffusion timesteps and a linear noise schedule ($\beta_t$ ranging from 0.00085 to 0.012). The trained teacher generates paired supervision samples: Gaussian noise inputs and corresponding target images. These pairs provide the training signal for the encoder-decoder pipeline.

\paragraph{Evaluation objective.}
The goal is to assess the empirical expressivity of the optical decoder by an ablation study: digital substitutes with the same inputs and outputs take its place. All decoders use the same encoder architecture and are evaluated on the same generation task. For each decoder, the complete encoder–decoder pipeline is trained end-to-end using the same training data and protocol.

\subsection{Pipeline Architecture}

\textbf{Digital encoder.} The encoder is a 3-layer class-conditional MLP with 3.3M parameters. The input concatenates flattened Gaussian noise ($32 \times 32 = 1024$ dimensions) with a learned class embedding. Three fully connected layers with LeakyReLU activations (slope 0.2) after the first two layers produce a 1025-dimensional output, which is split into a $32 \times 32$ phase pattern and a scalar scaling factor. A phase normalization coefficient $\alpha = 2.0$ maps the output to $[0, 2\pi]$.

\paragraph{Boundary of the module.}
All decoder modules (optical and digital) adhere to the same interface:
\begin{itemize}
    \item \textbf{Input:} Tensor of shape $(B, 1, 32, 32)$, where $B$ is the batch size.
    \item \textbf{Output:} Tensor of shape $(B, 1, 32, 32)$.
\end{itemize}
This $32 \times 32$ resolution matches the interface between the digital encoder and the loss computation stage. For the optical system, the $32 \times 32$ pattern is upsampled $10\times$ to $320 \times 320$ for the SLM, and the captured $320 \times 320$ intensity is downsampled $10\times$ back to $32 \times 32$.
Digital substitutes preserve these tensor dimensions but are not required to reproduce the optical hardware's physical input–output constraints; the comparison is based on task performance rather than pointwise functional equivalence.

\subsection{Decoder Architectures}

\subsubsection{Optical Decoder (OGM)}

The optical decoder is a Diffractive Deep Neural Network (D$^2$NN) with a single trainable phase mask:
\begin{itemize}
    \item \textbf{Phase mask:} $400 \times 400$ pixels, yielding 160K trainable phase pixels (``diffractive features'' in the original work), which are not parameters in the ML sense.
    \item \textbf{Propagation model:} Angular Spectrum Method at wavelength $\lambda = 520$ nm with a 5 cm propagation distance.
    \item \textbf{Readout:} Intensity measurement (magnitude-squared of the optical field).
\end{itemize}
The combination of phase encoding at the input and intensity detection at the output induces a nonlinear real-to-real mapping, though the optical propagation itself is linear.

We report two configurations:
\begin{itemize}
    \item \textbf{Optical (simulated):} The decoder is simulated using the code provided by the authors in \cite{chen2025optical}.
    \item \textbf{Optical (reported):} The experimental FID reported in~\cite{chen2025optical} (131.08), measured on the physical hardware, with its imperfections, and computed on $1{,}000$ generated images; our FIDs use $5{,}000$ samples (below), so the two are not strictly comparable.
\end{itemize}

\subsubsection{Linear Substitute}
A single fully-connected layer $1024 \to 1024$ with no nonlinearity, applied to the flattened $32\times 32$ input and reshaped back; $1.05$M parameters.

\subsubsection{MLP Substitutes}
Two depth families:
\begin{itemize}[leftmargin=1.5em, itemsep=0.15em]
    \item \textbf{MLP-2}: $\mathrm{Linear}(1024, h) \to \mathrm{LeakyReLU}(0.2) \to \mathrm{Linear}(h, 1024)$.
    \item \textbf{MLP-3}: $\mathrm{Linear}(1024, h) \to \mathrm{LeakyReLU}(0.2) \to \mathrm{Linear}(h, h) \to \mathrm{LeakyReLU}(0.2) \to \mathrm{Linear}(h, 1024)$.
\end{itemize}
We sweep the hidden dimension $h \in \{64, 128, 256, 512$$\}$ for both families, yielding $8$ MLP configurations.

\subsubsection{CNN Substitutes}
Two CNN variants with $31$K and $75$K parameters, all preserving spatial resolution at $32 \times 32$:
\begin{itemize}[leftmargin=1.5em, itemsep=0.15em]
    \item \textbf{CNN ($31$K)}: $\mathrm{Conv}(1, 128, 11{\times}11) \to \mathrm{BN} \to \mathrm{LeakyReLU} \to \mathrm{Conv}(128, 1, 11{\times}11)$.
    \item \textbf{CNN ($75$K)}: a $5$-layer CNN with $3{\times}3$ kernels and channel widths $1\to 32 \to 64 \to 64 \to 32 \to 1$, BatchNorm and LeakyReLU between layers.

\end{itemize}
We do not match the kernel size, depth, or channel width to any optical-module property: substitute architecture is selected from standard ML choices, and the only criterion is whether a substitute trained under standard ML practice matches or exceeds the optical module's reported task performance with the same inputs and outputs.

\subsection{Training Protocol}

\subsubsection{Loss Function}
All decoders are trained by knowledge distillation from a pretrained DDPM teacher~\cite{ho2020denoising}: given a teacher-generated paired sample $(z, y)$, where $z \sim \mathcal{N}(0, I)$ is a Gaussian noise tensor and $y$ is the teacher's denoised output, the student decoder $f$ is trained to minimize the pixel-wise mean-squared error
\begin{equation}
    \mathcal{L} = \mathbb{E}_{(z, y)} \big\| f(z) - y \big\|_2^2.
\end{equation}
Teacher samples are precomputed and cached, so every substitute consumes identical training pairs.

\subsubsection{Single Uniform Configuration}
A central design choice of our case study is to train digital substitutes under one standard training configuration, shared with the simulated optical module---neither deliberately under-tuned nor extensively re-optimized per architecture. We therefore use one fixed configuration for the simulated optical module and \emph{all} digital substitutes (Linear, MLP-2 sweep, MLP-3 sweep, all CNN variants):
\begin{itemize}[leftmargin=1.5em, itemsep=0.15em]
    \item Optimizer: AdamW~\cite{loshchilov2019decoupled} with $\beta_1{=}0.5$, $\beta_2{=}0.999$, weight decay $0$ (matching the optimizer setting in~\cite{chen2025optical}).
    \item Learning rate: $2\times 10^{-4}$ on all trainable digital parameters; for the simulated optical module only, the diffractive mask is additionally optimized at $10^{-2}$.
    \item Schedule: cosine annealing with a $10$-epoch linear warmup, over $150$ epochs total.
    \item Batch size: $100$.
    \item No exponential moving average; no early stopping.
\end{itemize}
The only architecture-specific element is the dedicated learning rate for the diffractive mask in the simulated optical module, which has no analogue in any digital substitute; no other hyperparameter is tuned per architecture or per capacity setting. We deliberately do not sweep learning rate or optimizer choice for the digital substitutes: a per-architecture sweep would (i) make the protocol architecture-specific and harder to reproduce, and (ii) grant digital substitutes a tuning budget that the optical module does not enjoy. The single-configuration choice keeps the ablation study on the safe side: it gives the substitutes no advantage that the optical module does not have (Sec.~\ref{sec:fairness}).

\subsubsection{Random Seeds and Reporting}
Each architecture/capacity setting is trained from scratch with $4$ independent seeds $\{42, 123, 456, 789\}$, controlling Python, NumPy, and PyTorch randomness. We report mean $\pm$~standard deviation FID at epoch~$150$ (final epoch). Across the simulated optical module, Linear, the MLP capacity sweep, and the two CNN variants, the reported sweep comprises $48$ individual training runs.

\subsection{Evaluation}

\paragraph{Sample generation.}
For each trained model and each seed, we generate $5{,}000$ samples by drawing $\mathbf{z} \sim \mathcal{N}(0, I)$, conditioning on uniformly sampled class labels $c \in \{0, \dots, 9\}$, and passing through the encoder--decoder pipeline.

\paragraph{Metrics.}
We report the Fr\'echet Inception Distance (FID)~\cite{heusel2017gans} between the $5{,}000$ generated samples and the MNIST test set, evaluated at epoch~$150$ (final epoch). We use 5,000 generated samples to obtain more stable FID estimates across runs~\cite{chong2020effectively}. FID is the same primary metric used in~\cite{chen2025optical}.

\paragraph{Compute.}
All experiments were conducted on NVIDIA A100 ($40$\,GB) and L$40$S ($48$\,GB) GPUs with mixed-precision training (FP$16$) where applicable. A single substitute configuration takes approximately $30$--$60$ minutes; the full $48$-run sweep takes up to about $50$ GPU-hours.

\subsection{Digital cost of the matched substitutes}
The matched MLP-3 ($h{=}128$) and MLP-2 ($h{=}256$) need about $0.56$ and $1.05$ million floating-point operations (FLOPs) per $32\times32$ image. Simulating the light is a natural first estimate of what a digital decoder would cost: in the authors' response in the published peer review file of Ref.~\cite{chen2025optical}, such a simulation ``will need 58.24 million digital FLOPs''. By comparison, a standard module that does the same job in the same place needs roughly $50$--$100\times$ fewer.

\begin{figure*}[t]
\centering
\includegraphics[width=.7\textwidth]{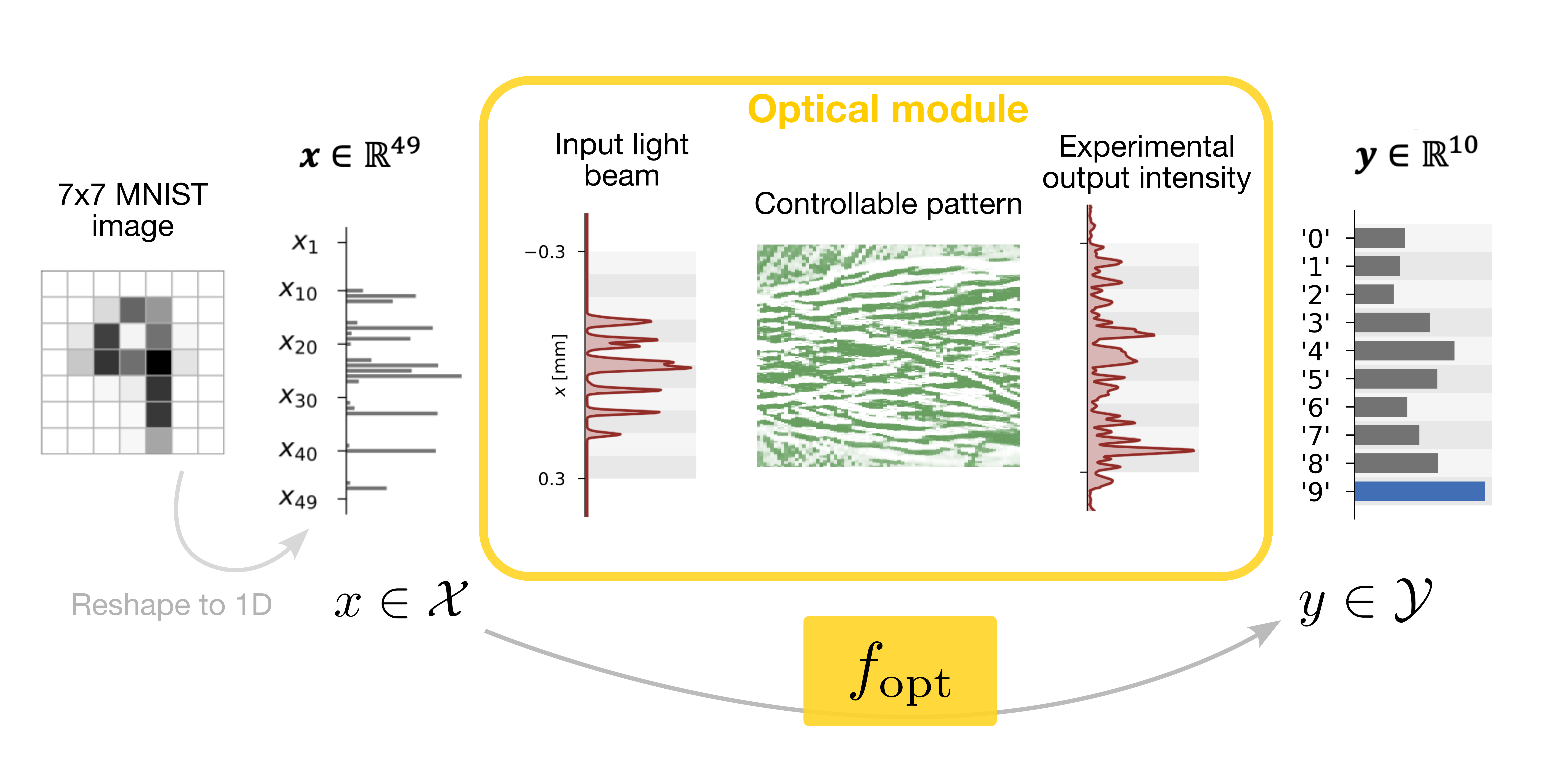}
\caption{Schematic of the 2D programmable waveguide pipeline for MNIST classification. A $7\times7$ MNIST image is flattened into an input vector $u \in \mathcal{X}$ (dimension 49), with each element encoded as the amplitude of a spatial mode. The optical module $f_{\mathrm{opt}}$ is a 2D programmable waveguide where spatially-varying refractive index modulation ($\Delta n$, shown in green) controls light propagation. The output $y \in \mathcal{Y}$ is obtained by binning the output intensity distribution into 10 spatial modes corresponding to digit classes `0'--`9'. Classification is determined by the channel with maximum intensity (here, `9'). Figure adapted from \cite{onodera2025arbitrary}.}
\label{fig:casea}
\end{figure*}

\section{Additional case study: 2D programmable waveguide}
\label{sec:caseA}
Here we demonstrate an operation-level example (Section~\ref{sec:background}) where the boundary and the ablation study are straightforward.
The 2D programmable waveguide \cite{wu2023lithography,onodera2025arbitrary} is a promising way to achieve arbitrary linear operations on the light field in integrated photonics. Its $\sqrt{N}$ depth scaling with respect to the number of modes $N$ \cite{onodera2025arbitrary} makes it more compelling than conventional integrated approaches such as MZI meshes \cite{shen2017deep}, which suffer from scalability limitations.

\paragraph{Task and interface.}
We consider the MNIST classification setup reported in \cite{onodera2025arbitrary}. Inputs are downsampled $7\times 7$ grayscale images, yielding a 49-dimensional nonnegative vector. Each element is encoded as a modal amplitude. Light propagates through the programmed waveguide, and the output distribution is binned into 10 channels corresponding to digit classes. Classification follows standard argmax over measured intensities (Figure~\ref{fig:casea}).

Following Section~\ref{sec:boundary}, the optical module maps a 49-dimensional real control vector (modal amplitudes) to a 10-dimensional real measurement vector (binned intensities). Internal physical details (scattering, drift, detector noise) are absorbed into the realized $f_{\mathrm{opt}}$.

\paragraph{Results of the ablation study.}
We replace $f_{\mathrm{opt}}$ with digital substitutes respecting the same $49 \to 10$ interface. A linear layer achieves 90\% test accuracy, compared to the reported 86\% for the optical system. For this task and interface, the optical module's empirical expressivity is comparable to a linear layer.
The authors themselves characterize the device by the operations of ``the effective 1-layer neural network'' it implements ($49\times10 = 490$ multiply-accumulate operations) rather than by those of the full optical transformation \cite{onodera2025arbitrary}, which is the card's logic applied at the operation level.

\begin{expressivitycard}{\; 2D programmable waveguide}
\small
\begin{enumerate}[leftmargin=1.5em, itemsep=0.35em, label=\textbf{\arabic*.}]
    \item \textbf{Input:} $u \in \mathbb{R}^{49}$ (nonnegative amplitudes; any normalization conventions as reported).
    \item \textbf{Output:} $y \in \mathbb{R}_{+}^{10}$ (binned intensities).
    \item \textbf{Empirical expressivity:} $\lesssim$ a linear layer.
\end{enumerate}
\end{expressivitycard}

This case illustrates that for ``operation-level'' ONNs, Expressivity Cards are close to what many papers already communicate implicitly. The card makes the ML-facing module definition explicit and makes the ablation study straightforward.

\section{Additional discussion}
\label{sec:discussion}

This section consolidates issues that commonly arise when optics and ML communities talk past each other. We keep the proposal itself simple (Expressivity Cards), but clarify what the card can and cannot capture and why it supports clearer communication.

\subsection{Keeping the card minimal: what belongs in the card}

The Expressivity Card is intended as a minimal add-on: a compact translation layer for each reported task. It should not become a device datasheet.

The card should contain only what an ML reader needs to (i) understand what real tensors flow into and out of the optical hardware, and (ii) interpret the module's computational capacity via an empirical expressivity report.

\subsection{Why empirical expressivity enables fair system-level claims}
\label{sec:discussion_resources}

A major motivation for ONNs is system-level advantages, such as low latency, low energy, high throughput, or favorable scaling. Such claims are meaningful only if the compared systems perform comparably on the same task under comparable I/O constraints, here the task the ONN authors chose to demonstrate.

This is where empirical expressivity is especially useful. The matching digital substitutes provide natural performance-matched digital references. This reference can then be used to state system-level claims, such as energy per inference, latency, or throughput, counted from the substitute's operations, in a way that ML readers can interpret; Sec.~\ref{sec:systemclaims} describes how.

This does not invalidate physics-native reporting (e.g., device-level throughput of modulator updates). Rather, it clarifies what those numbers should be compared to when the claim is ``this helps ML''.

\subsection{Task, architecture and optimization}

Empirical expressivity is inevitably entangled with the task and with the full pipeline in which the optical module is embedded. This is not unique to ONNs: in any empirical ML paper, reported performance reflects the dataset, architecture, and training procedure that the authors choose.

Expressivity Cards do not attempt to judge whether these choices are ``optimal.'' They are an add-on to the results the paper \emph{already} reports: given the authors' declared task, surrounding architecture, and evaluation metric, the card asks for (i) an explicit boundary/interface and (ii) a small set of standard digital substitutes trained with standard ML practice. The results should be interpreted as ``how much standard digital capacity is needed to match this reported configuration,'' not as a universal measure of the physical platform.

When training efficiency or limited-tuning claims are central to a paper, authors can optionally report the tuning protocol for the substitutes as part of the broader system discussion. Otherwise, standard training defaults are typically sufficient for the ablation study the card asks for.

\subsection{Noise, precision, and the simulation--experiment gap}
\label{subsec:noise}

Optical hardware is analog and subject to noise, drift, and calibration error, which is also why its system-level claims, such as speed or energy, are best stated per inference on a task (Sec.~\ref{sec:motivation}). ML readers are often used to deterministic forward passes on GPUs; experimental variability can therefore be surprising. Importantly, this gap is not accidental: many ONN designs are proposed and optimized under an idealized forward model, while the realized experiment introduces additional imperfections. In practice, the best achievable performance in simulation is often an \emph{upper bound} on what an experimental prototype can demonstrate without additional calibration and control. This simulation--experiment comparison is already common in ONN papers and provides a useful signal of how faithfully the intended mechanism translates to hardware.

A function-oriented view does not ignore these uncertainties. It treats noise and variability as part of the realized input--output behavior and recommends making the evaluation context explicit. We therefore suggest two complementary reporting modes:
\begin{itemize}
\item \textbf{Simulation card (idealized).} Report the card, boundary and ablation study, under the simulation model used for training/analysis. This reflects the intended mechanism and provides a useful ``best-case'' reference for computational capability.
\item \textbf{Hardware card (realized).} Report the same fields for the experimental system, including the evaluation protocol and any calibration procedure. This reflects the realized behavior of the physical module under noise, drift, and measurement nonidealities.
\end{itemize}
Comparing these two cards is informative: it separates limitations of the proposed mechanism (simulation) from limitations of implementation (hardware), without requiring ML readers to reason in device physics.
Hysteresis and other history-dependent behavior are treated the same way. The card does not require a closed-form model of the optical mapping, nor a substitute that reproduces it: the substitute is trained on the task, and the optical module's measured performance already includes these effects.

\paragraph{Stochasticity and randomness.}
Some ONN systems intentionally exploit stochasticity (e.g., random scattering, shot noise, or other noise sources) as part of generation or sampling. For \emph{classical} ML tasks, stochasticity can also be modeled digitally via standard random number generation---as routinely done in GANs, diffusion models, and stochastic regularization. Therefore, if randomness is part of the \emph{function} being implemented, it can be incorporated into digital substitutes as well (e.g., by sampling noise in the substitute or by adding controlled perturbations at the interface). If a paper's claim is that physical noise is ``more intrinsic'' or ``free,'' that is a system-level motivation (hardware simplicity, latency, energy) and should be evaluated as such; it does not remove the need to report the implemented mapping and its task performance.

\subsection{The optimized optical module beats an unoptimized baseline; isn't that sufficient?}
\label{sec:optbaseline}

These comparisons are valuable and should remain standard. A common pattern in ONN papers is to compare an optimized optical configuration against a non-optimized or ``naive'' physical baseline---for example, using a flat phase mask (no modulation), random phase masks, bypassing the optical module, or free-space propagation without any programmed diffraction. Such results show that the physical degrees of freedom are being used meaningfully and that the learned configuration is not arbitrary. They are ablation studies too, of the optics' own design choices; Appendix~\ref{app:census-comparisons} discusses removing the optics.

However, this style of baseline typically answers a different question: \emph{``did optimization make the device do something nontrivial?''} It does not answer the ML-facing question: \emph{``how much computational capacity does this module provide, compared to standard digital substitutes under the same interface and metric?''} Expressivity Cards complement (not replace) such optical baselines by adding the ablation study ML readers look for first: standard digital substitutes in the optical module's place. This makes the module's contribution interpretable to ML readers while preserving the value of optics-native ablations.

\subsection{Why don't we have case studies for more ONN papers?}
\label{sec:morecases}

Applying the Expressivity Card retrospectively requires reproducing each paper's training and evaluation pipeline, then replacing the optical module with standard digital substitutes. For many ONN systems we considered, especially the task-level ones, the released code, data, or implementation details were not sufficient for such controlled replication. Even when the architecture is clear, retraining an optical pipeline with matched substitutes can be non-trivial.

This limitation is part of the motivation for the EC. The EC is not intended as a retrospective audit of existing ONN papers; it is a lightweight reporting norm for future papers to include alongside their original evaluations. If authors report this ablation study at publication time, the accessibility problem is addressed at the source.

We therefore apply the EC where replication was feasible. We are grateful that the OGM authors made their code available, enabling our replication. A broader quantitative survey of expressivity would be valuable, but it would require exactly the kind of replication effort that is currently limited by incomplete code and data availability. Reporting practice, in contrast, can be examined without replication; Appendix~\ref{app:census} does so for a set of recent task-level papers.

\subsection{Should the digital substitutes match parameter count, FLOPs, etc.?}
\label{sec:paramcount}

Following the function-oriented principle, we do not recommend matching any of these quantities. Direct comparison of parameter counts or FLOPs between optical and digital systems is a common source of confusion, and is usually impossible given the fundamental differences between optical modules and digital conventions.

On the contrary, one of the primary purposes of the EC is to provide the ``anchors'' to these quantities in a way that is interpretable to ML readers: once a performance-matched digital substitute is identified, its parameter count and FLOPs give a concrete, ML-native reference point for evaluating system-level claims (such as what kind of digital system the optical module could replace; Sec.~\ref{sec:systemclaims}).

In practice, reporting the quantities like parameter counts for the digital substitutes is straightforward and encouraged---these are ML-native quantities that give readers immediate intuition. Reporting them strengthens both the empirical insights and the reproducibility of the Expressivity Card.
\section{Industry Investment in Optical Computing}
\label{app:industry}

The optical computing industry sector has attracted substantial venture capital investment in recent years, reflecting widespread belief in the technology's potential to address AI infrastructure challenges. Table~\ref{tab:onn-funding} summarizes funding for both major as well as early-stage optical computing startups as of May 2026.

\begin{table*}[t]
\centering
\caption{Venture funding and valuations for optical computing companies (as of May 2026). Source numbers refer to the list below the table.}
\label{tab:onn-funding}
\small
\begin{tabularx}{\linewidth}{p{0.17\linewidth} p{0.2\linewidth} r r l}
\toprule
\textbf{Company} & \textbf{Focus} & \textbf{Total Raised} & \textbf{Valuation} & \textbf{Latest Event} \\
\midrule
\multicolumn{5}{l}{\textit{Tier 1: Public / Acquired}} \\
\midrule
Lightelligence$^1$ & Computing/Interconnect & \$765M+ & \$10.4B (mkt cap) & IPO (Apr 2026) \\
Celestial AI$^2$ & Interconnect & \$520M & Acq. \$5.5B & Acquired (Dec 2025) \\
\midrule
\multicolumn{5}{l}{\textit{Tier 2: Unicorns ($>$\$1B valuation)}} \\
\midrule
Ayar Labs$^3$ & Interconnect & \$870M & \$3.75B & Series E (Mar 2026) \\
Lightmatter$^4$ & Computing/Interconnect & \$850M & \$4.4B & Series D (Oct 2024) \\
OLIX Computing$^5$ & Computing & \$250M & $>$\$1B & Series A (Feb 2026) \\
\midrule
\multicolumn{5}{l}{\textit{Tier 3: Growth Stage (\$100M+ raised)}} \\
\midrule
Xanadu$^6$ & Quantum/Computing & \$287M & --- & Series B (May 2021) \\
Neurophos$^7$ & Computing & \$120M & --- & Series A (Jan 2026) \\
Luminous Computing$^8$ & Computing & \$114M & --- & Series A (Mar 2022) \\
\midrule
\multicolumn{5}{l}{\textit{Tier 4: Mid Stage (\$20M--\$100M raised)}} \\
\midrule
OpenLight$^9$ & Interconnect & \$84M & --- & Series A-1 (Apr 2026) \\
Q.ANT$^{10}$ & Computing & \$80M & --- & Series A (Oct 2025) \\
Optalysys$^{11}$ & Computing & \$29M & --- & Series A ext. (Jan 2026) \\
Arago$^{12}$ & Computing & \$26M & --- & Seed (Jul 2025) \\
iPronics$^{13}$ & Interconnect & \texteuro 20M & --- & Series A (Jan 2025) \\
NcodiN$^{14}$ & Interconnect & \texteuro 16M & --- & Seed (Nov 2025) \\
\midrule
\multicolumn{5}{l}{\textit{Tier 5: Early Stage ($<$\$20M raised)}} \\
\midrule
Opticore$^{15}$ & Computing & \$14.5M & --- & Seed ext. (Sep 2025) \\
LightSolver$^{16}$ & Computing & \texteuro 12.5M & --- & EIC Grant (Aug 2024) \\
Lumai$^{17}$ & Computing & \$10M & --- & Series A (Apr 2025) \\
Volantis$^{18}$ & Computing/Interconnect & \$9M & --- & Seed (Jun 2025) \\
Akhetonics$^{19}$ & Computing & \texteuro 8.3M & --- & Seed (Nov 2024) \\
Cognifiber$^{20}$ & Computing & \$8.5M & --- & Seed (2022) \\
\midrule
\textbf{Total} & & \textbf{$>$\$4B} & \textbf{$>$\$25B} & \\
\bottomrule
\end{tabularx}

\vspace{0.5em}
\raggedright
\scriptsize
\textbf{Sources:}
$^1$\href{https://www.bloomberg.com/news/articles/2026-04-27/optical-computing-firm-lightelligence-set-for-hk-debut-after-ipo}{Optical Computing Firm Lightelligence Set for HK Debut After IPO}, Bloomberg, Apr 2026.
$^2$\href{https://optics.org/news/16/11/47}{Marvell looks to scale up optical links with \$5BN-plus Celestial AI deal}, Optics.org, Dec 2025.
$^3$\href{https://ayarlabs.com/news/ayar-labs-closes-500m-series-e-accelerates-volume-production-of-co-packaged-optics/}{Ayar Labs Closes \$500M Series E}, Ayar Labs, Mar 2026.
$^4$\href{https://www.businesswire.com/news/home/20241016498931/en/}{Lightmatter Raises \$400M Series D; Quadruples Valuation to \$4.4B}, BusinessWire, Oct 2024.
$^5$\href{https://techfundingnews.com/olix-220m-unicorn-photonic-ai-chips-inference/}{Olix nabs \$220M for photonic AI inference chips}, TechFundingNews, Feb 2026.
$^6$\href{https://thequantuminsider.com/2026/03/24/11-companies-lighting-up-the-quantum-photonics-sector/}{11 Companies Lighting Up the Quantum Photonics Sector}, The Quantum Insider, Mar 2026.
$^7$\href{https://www.prnewswire.com/news-releases/neurophos-secures-110-million-series-a-to-launch-exaflop-scale-photonic-ai-chips-302667517.html}{Neurophos Secures \$110M Series A}, PR Newswire, Jan 2026.
$^8$\href{https://www.theregister.com/2022/03/05/luminous-ai-supercomputer-photonics/}{Photonics startup Luminous Computing in \$105m series-A round}, The Register, Mar 2022.
$^9$\href{https://www.businesswire.com/news/home/20260428377852/en/}{OpenLight Secures \$50M in Series A-1 Funding}, BusinessWire, Apr 2026.
$^{10}$\href{https://qant.com/press-releases/duquesne-family-office-invests-in-q-ant-to-drive-sustainable-photonic-ai-infrastructure/}{Duquesne Family Office Invests in Q.ANT}, Q.ANT, Oct 2025.
$^{11}$\href{https://optics.org/news/17/1/32}{Optalysys eyes US expansion with \pounds 23M}, Optics.org, Jan 2025.
$^{12}$\href{https://www.eu-startups.com/2025/07/french-deeptech-company-arago-raises-e22-1-million-to-slash-ai-energy-consumption-with-innovative-photonic-chip/}{French DeepTech company Arago raises EUR 22.1M}, EU-Startups, Jul 2025.
$^{13}$\href{https://www.globenewswire.com/news-release/2025/01/28/3016683/0/en/iPRONICS-Raises-20M-in-Series-A-to-Advance-Optical-Networking-for-Next-Gen-AI-Data-Centers.html}{iPronics Raises EUR 20M in Series A}, GlobeNewswire, Jan 2025.
$^{14}$\href{https://www.businesswire.com/news/home/20251119479101/en/}{NcodiN Secures EUR 16M Seed Round}, BusinessWire, Nov 2025.
$^{15}$\href{https://www.accessnewswire.com/newsroom/en/computers-technology-and-internet/opticore-raises-14-million-to-advance-photonic-computing-chips-10-1071208}{Opticore Raises \$14M}, ACCESS Newswire, Sep 2025.
$^{16}$\href{https://www.calcalistech.com/ctechnews/article/b1j8vqe50}{LightSolver awarded EUR 12.5M from EIC}, Calcalist Tech, Aug 2024.
$^{17}$\href{https://optics.org/news/16/4/7}{Photonic computing startup Lumai lands \$10M}, Optics.org, Apr 2025.
$^{18}$\href{https://www.businesswire.com/news/home/20250611870483/en/}{Volantis Raises \$9M in Seed Round}, BusinessWire, Jun 2025.
$^{19}$\href{https://techcrunch.com/2024/11/28/akhetonics-gets-fresh-funding-for-a-contrarian-bet-on-all-optical-chips/}{Akhetonics gets fresh funding}, TechCrunch, Nov 2024.
$^{20}$\href{https://www.techinsights.com/blog/cognifiber-targets-purely-photonic-ai}{Cognifiber Targets Purely Photonic AI}, TechInsights, 2022.
\end{table*}

\paragraph{Public market validation.} Lightelligence's Hong Kong IPO on April 28, 2026---the first pure-play photonic computing company to go public---delivered a 383.6\% first-day gain and a market capitalization of \$10.4 billion. The retail tranche was oversubscribed 5,785$\times$, with cornerstone investors including Alibaba, Lenovo, Baillie Gifford, and BlackRock. This debut signals that capital markets now view photonic computing as credible AI infrastructure rather than speculative research.

\paragraph{Strategic acquisitions and investments.} Marvell's acquisition of Celestial AI (announced December 2025, valued at \$3.25 billion upfront and up to \$5.5 billion with earnouts) and AMD's acquisition of Enosemi (May 2025) demonstrate that semiconductor incumbents are actively integrating optical interconnect capabilities. These acquisitions, combined with Nvidia's and AMD's strategic investments in Ayar Labs' \$500 million Series E (March 2026), indicate strong conviction in optical technology for AI infrastructure. High-profile investors---including Bill Gates (Luminous), Stanley Druckenmiller (Q.ANT), and Sam Altman (Volantis)---have made early-stage bets across the sector.

\paragraph{Evaluation gap.} Despite this substantial capital deployment---now exceeding \$4 billion across tracked companies, with three unicorns and one public company valued above \$10 billion---we are unaware of any standardized methodology for comparing the computational expressivity of these systems against conventional digital alternatives. The reporting practice proposed in this paper aims both to address this gap and to highlight its importance.

\end{document}